\documentclass[prd,twocolumn,amsmath,amssymb,unsortedaddress,nofootinbib]{revtex4-1}
\usepackage{graphicx}
\def\pmb#1{\setbox0=\hbox{$#1$}%
  \kern-.025em\copy0\kern-\wd0
  \kern.05em\copy0\kern-\wd0
  \kern-.025em\raise.0433em\box0}
\def\parb{\pmb{\partial}}
\def\alt{\mathrel{\hbox{\rlap{\hbox{\lower4pt\hbox{$\sim$}}}\hbox{$<$}
}}}
\def\be{\begin{equation}}
\def\ee{\end{equation}  }
\def\bea{\begin{eqnarray}}
\def\eea{\end{eqnarray}  }

\begin{document}
\title{Spike behavior in the approach to spacetime singularities}

\author{David Garfinkle}
\email{garfinkl@oakland.edu}
\affiliation{Department of Physics, Oakland University, Rochester, MI 48309}
\affiliation{Michigan Center for Theoretical Physics, Randall Laboratory of Physics, University of Michigan, Ann Arbor, MI 48109-1120}
\author{Frans Pretorius}
\email{fpretori@princeton.edu}
\affiliation{Joseph Henry Laboratories, Princeton University, Princeton, NJ 08544}

\begin{abstract}
We perform numerical simulations of the approach to spacetime singularities.  The simulations are done with sufficient resolution to resolve the small scale features (known as spikes) that form in this process.  We find an analytical formula for the shape of the spikes and show that the spikes in the simulations are well described by this formula.
\end{abstract}

\maketitle

\section{Introduction}
\label{sec_intro}

Ever since the singularity theorem of Penrose~\cite{penrose}, it has been known that spacetime singularities are a generic feature of gravitational collapse.  However, Penrose's theorem gives very little information about the nature of these singularities, stating only that some light ray fails to be complete.  In order to obtain a better understanding of the nature of singularities, Belinskii, Lifschitz, and Khalatnikov~\cite{bkl} (collectively known as BKL) conjectured an analytic approximation in which near the singularity, terms in the field equations containing spatial derivatives were negligible compared to those containing time derivatives.  In order to test the correctness 
of the BKL conjecture, Berger and Moncrief~\cite{beverlyandvince1} performed numerical simulations of the approach to the singularity in Gowdy spacetimes.  The Gowdy spacetimes have two spatial Killing vectors, and thus form a rather specialized class of spacetimes, which can be thought of as a toy model for the general problem of gravitational  collapse.  Nonetheless, even in this special case Berger and Moncrief found a new and unexpected feature of singularities: as the singularity was approached the dynamics at almost all spatial points was in accord with the BKL conjecture; however, there were isolated points at which sharp features developed and became ever narrower the nearer one got to the singularity.  

These sharp features later became known as spikes.  The spikes represent a challenge for numerical simulations because an accurate numerical simulation requires that the spatial points that make up the numerical grid have sufficiently small separation to resolve all features.  For a fixed spatial resolution, an ever narrowing spatial feature, such as the spikes found in~\cite{beverlyandvince1} will eventually become too narrow to be resolved.  However, because the Gowdy spacetimes have two spatial Killing fields, numerical simulations of these spacetimes require only a single spatial dimension, and thus can be done with a very fine spatial resolution.  In~\cite{dgandbeverly} these fine scale numerical simulations were compared with an approximate analytical formula for the behavior of the spikes and were shown to match that formula. In~\cite{lim} a class of exact analytic solutions was
found for spikes in Gowdy spacetimes, and shown in~\cite{lim_et_al} to approach the late-time behavior
of numerical simulations of spike formation in $G_2$ spacetimes (a generalization
of the Gowdy model, but still with two spatial Killing vectors). Thus, despite the numerical challenges that they pose, spikes in Gowdy spacetimes are well understood.

The work of~\cite{beverlyandvince1} was generalized to the case of only one Killing field~\cite{beverlyandvince2,beverlyetal} and later (using a different numerical method based on the analytical work of~\cite{Uggla:2003fp}) to the case of no symmetry~\cite{dgprl}. However, the simulations of ~\cite{beverlyandvince2} and ~\cite{dgprl} did not have sufficient resolution to resolve the spikes.  One method to obtain better resolution is adaptive mesh refinement (AMR)~\cite{BO}, which detects when resolution is about to become insufficient and then adds extra spatial points where they are needed.  Indeed, AMR was used to resolve spikes in Gowdy spacetimes by Hern and Stewart~\cite{stewart}. However, though AMR is an effective method to use on Gowdy spacetimes, it is not so effective for the case of only one symmetry, or for the case of no symmetry.  This is because AMR works well when the features that it needs to resolve occur at isolated spatial points, while (as we will see later) spikes are features of co-dimension one: that is, spikes occur at isolated points in the case of two symmetries, along curves in the case of one symmetry, and  at surfaces in the case of no symmetry.  Thus, in the later two cases, the AMR would need to resolve too many regions and would quickly be overwhelmed.  
To obtain answers with adequte resolution in a reasonable amount of time thus requires that we parallelize the code; we use
the PAMR/AMRD~\cite{pamr_amrd} libraries to do this. Our highest resolution run used $112$ cores of the {\em Perseus} cluster
at Princeton, taking two days to complete.

In section \ref{equations}, we present the field equations used in our simulations.  These are the vacuum Einstein field equations expressed in terms of the scale invariant variables of~\cite{Uggla:2003fp}.  Section \ref{spikeformula} introduces a truncation of these equations obtained by applying the BKL approximation and derives an analytic formula for the spike from these truncated equations.  Subsections \ref{ephemeral} and \ref{earlylate} explore the implications of the approximations made in section \ref{spikeformula}.  Section \ref{1d} presents 1 dimensional (i.e. the case of two Killing fields) simulations of the equations of section \ref{equations} and the comparison of those results to the analytic formula of section \ref{spikeformula}.  Section \ref{2d} performs the same sort of simulations and comparison to analytic formula for the two dimensional (i.e. one Killing field) case.  Our conclusions are presented in section \ref{conclusions}. 

\section{Equations of motion}
\label{equations}

The method we use to evolve the vacuum Einstein equations is
the scale invariant tetrad method
of Uggla et al
~\cite{Uggla:2003fp}.
We use this method with constant mean curvature slicing as is done
in the simulations of
~\cite{dgaei} (or equivalently as is done in the cosmological simulations of ~\cite{ekpyro,ijjas_et_al} but with no scalar field matter).  More information on this type of method can be found in
~\cite{Uggla:2003fp,dgaei,ekpyro,ijjas_et_al}.   

The spacetime
is described in terms of a coordinate system ($t,{x^i}$) and a tetrad
(${{\bf e}_0},{{\bf e}_\alpha}$)  where both the spatial coordinate
index $i$ and
the spatial tetrad index $\alpha $ go from 1 to 3.  Choose
${\bf e}_0$ to be hypersurface orthogonal with the relation between
tetrad and coordinates of the form
${{\bf e}_0} = {N^{-1}}{\partial _t}$, and 
${{\bf e}_\alpha} =
{{e_\alpha }^i}{\partial _i},$
where $N$ is the lapse and the shift is chosen to be zero.
Choose the spatial frame $\{ {{\bf e}_\alpha} \}$ to be
Fermi propagated along the integral curves of ${\bf e}_0$.
The commutators of the tetrad components are decomposed as follows:
\begin{eqnarray}
[{{\bf e}_0},{{\bf e}_\alpha}] &=& {{\dot u}_\alpha}{{\bf e}_0}
-(H {{\delta _\alpha}^\beta}
+{{\sigma _\alpha}^\beta})
{{\bf e}_\beta}
\label{commute1}
\\
\left [ {{\bf e}_\alpha },{{\bf e}_\beta} \right ]  &=&
(2 {a_{[\alpha}}{{\delta _{\beta ]}}^\gamma}
+ {\epsilon _{\alpha \beta \delta }}{n^{\delta \gamma}}){{\bf 
e}_\gamma},
\label{commute2}
\end{eqnarray}
where $n^{\alpha \beta}$ is symmetric, and $\sigma ^{\alpha \beta}$ is
symmetric and trace free.
The scale invariant tetrad variables are defined by 
${\parb_0} \equiv 
{{\bf e}_0}/H$ and ${\parb_\alpha} \equiv {{\bf e}_\alpha}/H$
while scale invariant versions of the other gravitational 
variables are given by
\begin{equation}
\{ {{E_\alpha}^i}, {\Sigma _{\alpha \beta }}, {A^\alpha} ,
{N_{\alpha \beta }} \} \equiv \{ {{e_\alpha}^i} ,
{\sigma _{\alpha \beta }} , {a^\alpha}, {n_{\alpha \beta}} \} /H.
\end{equation}
Note that the relation between the scale invariant tetrad variables 
and
the coordinate derivatives is
\begin{eqnarray}
{\parb_0} &=& {{\cal N}^{-1}} {\partial _t}\label{d0_def}
\\
{\parb_\alpha} &=& {{E_\alpha}^i}{\partial _i},
\end{eqnarray}
where ${\cal N} = NH$ is the scale invariant lapse.  
The time coordinate $t$ is chosen so that 
\begin{equation}
{e^{-t}} = 3 H.
\label{timechoice}
\end{equation}
Here we have used the scale invariance of the physical system to make
both $t$ and $H$ dimensionless quantities.
Note that equation (\ref{timechoice}) 
means that the surfaces of constant time are constant mean 
curvature surfaces. Note also that the singularity is approached as
$t \to - \infty$.  

\noindent Due to equation (\ref{timechoice}) the scale invariant lapse satisfies 
an elliptic equation
\be
- {\parb ^\alpha}{\parb _\alpha} {\cal N} + 2 {A^\alpha} {\parb _\alpha} {\cal N} 
+ {\cal N} (3 + {\Sigma _{\alpha \beta}}{\Sigma ^{\alpha \beta}} ) = 
3 \label{Neqn}.
\ee
The gravitational quantities ${{E_\alpha}^i}, \, {A_\alpha}, 
{N^{\alpha \beta}}$ and $\Sigma _{\alpha \beta}$ satisfy the following 
evolution equations.
\bea
{\partial _t} {{E_\alpha}^i} &=& {{E_\alpha}^i} - {\cal N} 
({{E_\alpha}^i}
+ {{\Sigma _\alpha}^\beta}{{E_\beta}^i}),
\label{dtE}\\
{\partial _t} {A_\alpha} &=& {A_\alpha} + {\textstyle {\frac 1 2}} 
{{\Sigma _\alpha}^\beta}{\parb _\beta}{\cal N} -
{\parb _\alpha}{\cal N} \nonumber\\
&\ & + {\cal N} \left (
 {\textstyle {\frac 1 2}} {\parb _\beta}{{\Sigma _\alpha}^\beta}
- {A_\alpha} - {{\Sigma _\alpha}^\beta}{A_\beta} \right ),
\label{dtA}
\\
{\partial _t} {N^{\alpha \beta}} &=& {N^{\alpha \beta}} - 
{\epsilon ^{\gamma \delta ( \alpha}}{{\Sigma _\delta}^{\beta )}}
{\parb _\gamma} {\cal N} 
+ {\cal N} \Bigl ( - {N^{\alpha \beta}} \nonumber\\
&\ & 
+ 2 {{N^{(\alpha}}_\gamma}{\Sigma ^{\beta )\gamma}} - 
{\epsilon ^{\gamma \delta ( \alpha}} {\parb _\gamma} {{\Sigma 
_\delta}^{\beta )}} \Bigr )   
\label{dtN},
\\
{\partial _t} {\Sigma _{\alpha \beta}} &=& {\Sigma _{\alpha \beta}} 
+ {\parb _{<\alpha}}{\parb _{\beta >}} {\cal N} + {A_{<\alpha}}
{\parb _{\beta >}} {\cal N} \nonumber\\ 
&\ & 
+ {\epsilon _{\gamma \delta (\alpha}}
{{N_{\beta )}}^\delta}{\parb ^\gamma}{\cal N}
+ {\cal N} \Bigl [ - 3 {\Sigma _{\alpha \beta}} 
\nonumber
\\
&\ &- {\parb _{<\alpha}}
{A_{\beta >}}
- 2 {{N_{<\alpha}}^\gamma}{N_{\beta >\gamma}}
+ {{N^\gamma }_\gamma}{N_{< \alpha \beta >}} \nonumber\\
&\ & +  {\epsilon _{\gamma \delta (\alpha}} ( {\parb ^\gamma} {{N_{\beta
)}}^\delta}
- 2 {A^\gamma} {{N_{\beta )}}^\delta}) \Bigr ].
\label{dtSig}
\eea
Here parentheses around a pair of indices denote the symmetric part, while
angle brackets denote the symmetric trace-free part.

In addition, the variables are subject to the vanishing of the 
following 
constraint quantities
\bea
{{({{\cal C}_{\rm com}})}^{\lambda i}} &=& {\epsilon ^{\alpha \beta 
\lambda}}
[ {\parb _\alpha} {{E_\beta}^i} - {A_\alpha} {{E_\beta}^i} ] - 
{N^{\lambda \gamma}}{{E_\gamma}^i},
\label{constraintCOM}
\\
{{({{\cal C}_J})}^\gamma} &=& {\parb _\alpha}{N^{\alpha \gamma}} + 
{\epsilon ^{\alpha \beta \gamma}} {\partial _\alpha}{A_\beta}
- 2 {A_\alpha}{N^{\alpha \gamma}},
\label{constraintJ}
\\
{{({{\cal C}_C})}_\alpha} 
&=& {\parb _\beta}{{\Sigma _\alpha}^\beta}
 - 3 {{\Sigma _\alpha}^\beta}{A_\beta} - {\epsilon _{\alpha \beta 
\gamma}}
{N^{\beta \delta}}{{\Sigma _\delta}^\gamma},
\label{constraintC}
\\
{{\cal C}_G} &=& 1 + {\textstyle {\frac 2 3}}{\parb _\alpha}{A^\alpha}
- {A^\alpha}{A_\alpha}
- {\textstyle {\frac 1 6}}{N^{\alpha \beta}}{N_{\alpha \beta}} \nonumber\\
&\ &
+ {\textstyle {\frac 1 {12}}}{{({{N^\gamma}_\gamma})}^2} 
- {\textstyle {\frac 1 6}}{\Sigma ^{\alpha \beta}}{\Sigma_{\alpha 
\beta}}.
\label{constraintG}
\eea

Initial data is chosen to satisfy the constraints of eqns. (\ref{constraintCOM}-\ref{constraintG}), which are then preserved (to within numerical truncation error) under evolution.  The data are evolved using the evolution equations, eqns. (\ref{dtE}-\ref{dtSig}), where to obtain a hyperbolic system a multiple of eqn. (\ref{constraintC}) is added to the right hand side of 
eqn. (\ref{dtA})~\cite{dgandcg}.

\section{Universal spike behavior}
\label{spikeformula}

We now derive an analytic approximation for the shape of the spikes.  The BKL conjecture for the system in this form says that sufficiently close to the singularity, $A_\alpha$ and ${{E_\alpha}^i}$ are small enough to be neglected.  Note that all spatial derivatives occur in the equations of motion in the form ${\parb _\alpha} = {{E_\alpha}^i}{\partial _i}$, so it follows that all these terms can also be neglected.  Subject to this approximation, we find the following: eqn. (\ref{Neqn}) becomes  
\be
{{\cal N}^{-1}} = 1 + {\textstyle {\frac 1 3}}{\Sigma _{\alpha \beta}}{\Sigma ^{\alpha \beta}}.
\label{bklNeqn}
\ee
Eqns. (\ref{dtE})and (\ref{dtA}) are automatically satisfied.  Eqns. (\ref{dtN}) and (\ref{dtSig}) become
\bea
{\partial _t} {N^{\alpha \beta}} &=& {N^{\alpha \beta}} 
+ {\cal N} \Bigl ( - {N^{\alpha \beta}}
+ 2 {{N^{(\alpha}}_\gamma}{\Sigma ^{\beta )\gamma}} \Bigr ),
\label{bkldtN}
\\
{\partial _t} {\Sigma _{\alpha \beta}} &=& {\Sigma _{\alpha \beta}} + {\cal N} \Bigl [ - 3 {\Sigma _{\alpha \beta}}
- 2 {{N_{<\alpha}}^\gamma}{N_{\beta >\gamma}}\nonumber\\
&\ &
+ {{N^\gamma }_\gamma}{N_{< \alpha \beta >}} \Bigr ].
\label{bkldtSig}
\eea
Eqns. (\ref{constraintCOM}) and (\ref{constraintJ}) are automatically satisfies, while eqns. (\ref{constraintC}) 
and (\ref{constraintG}) become
\bea
{\epsilon _{\alpha \beta \gamma}} {N^{\beta \delta}}{{\Sigma _\delta}^\gamma} = 0,
\label{bklconstraintC}
\\
1 - {\textstyle {\frac 1 6}}{N^{\alpha \beta}}{N_{\alpha \beta}} 
+ {\textstyle {\frac 1 {12}}}{{({{N^\gamma}_\gamma})}^2} 
- {\textstyle {\frac 1 6}}{\Sigma ^{\alpha \beta}}{\Sigma_{\alpha \beta}} = 0. 
\label{bklconstraintG}
\eea

We begin at an initial time close enough to the singularity that the conditions of the BKL conjecture are satisfied and follow the behavior through one bounce.
Eqn. (\ref{bklconstraintC}) implies that the matrices 
${\Sigma ^\alpha}_\beta$ and ${N^\alpha}_\beta$ commute and therefore have a common basis of eigenvectors.  It then follows from eqns. (\ref{bkldtN}) and (\ref{bkldtSig}) that the eigenvectors are constant in time (see Appendix B of ~\cite{ijjas_et_al} for more details); so all that we need to do is find the time dependence of the eigenvalues.  Denote the eigenvalues of ${\Sigma ^\alpha}_\beta$ by 
${\Sigma_1}, \, {\Sigma_2}$ and $\Sigma_3$
with ${\Sigma_1}\le {\Sigma_2} \le {\Sigma_3}$ at the initial time.  Let $N_1$ be the eigenvalue of ${N^\alpha}_\beta$
corresponding to the eigenvector of ${\Sigma ^\alpha}_\beta$ that has eigenvalue $\Sigma_1$, and correspondingly for 
$N_2$ and $N_3$.  We assume that at the initial time ${N_1}, \, {N_2}$ and $N_3$ are all negligibly small.  Then it follows
from eqn. (\ref{bkldtN}) that $N_1$ grows in magnitude during the bounce process, but that $N_2$ and $N_3$ decrease in magnitude and therefore remain negligible.  
We then find from eqns. (\ref{bklNeqn}) and (\ref{bklconstraintG}) that 
\be
{{\cal N}^{-1}} = 3 - {\textstyle {\frac 1 6}} {{({N_1})}^2}.
\label{bklNeqn2}
\ee
Using eqn. (\ref{bklNeqn2}) in eqns. (\ref{bkldtN}-\ref{bkldtSig}) we then obtain
\bea
{\partial _t} {\Sigma _1} = (1 - 3 {\cal N}) ({\Sigma _1} + 4),
\label{dtSig1}
\\
{\partial _t} {\Sigma _2} = (1 - 3 {\cal N}) ({\Sigma _2} -2),
\label{dtSig2}
\\
{\partial _t} {\Sigma _3} = (1 - 3 {\cal N}) ({\Sigma _3} -2),
\label{dtSig3}
\\
{\partial _t} {N_1} = {N_1} ( 1 + {\cal N} ( -1 + 2 {\Sigma _1} )).
\label{dtN1}
\eea
Now define the quantity $Z$ by
\be
Z \equiv {\Sigma _1} + 4.
\label{Zdef}
\ee
Then it follows from eqns. (\ref{dtSig1}-\ref{dtSig3}) and the fact that $\Sigma_{\alpha \beta}$ is trace-free that there is a constant $b$ such that 
\bea
{\Sigma_2} = 2 + (b-1) Z,
\label{Sig2}
\\
{\Sigma_3} = 2 - b Z.
\label{Sig3}
\eea
It then follows from eqns. (\ref{bklNeqn}) and (\ref{Zdef}-\ref{Sig3}) that 
\be
{{\cal N}^{-1}} - 3 = {\textstyle {\frac 2 3}} ( {b^2} - b + 1) {Z^2} - 4 Z + 6.
\label{Zquad1}
\ee
Thus ${{\cal N}^{-1}} - 3$ is a quadratic in $Z$.  Let $Z_+$ and $Z_-$ be the
roots of this quadratic.  Then we have
\be
{Z_\pm} = {\frac {3 \pm 3 {\sqrt {b(1-b)}}} {{b^2} - b + 1}}.
\label{Zpm}
\ee
Using eqn. (\ref{Zpm}) in eqn. (\ref{Zquad1}) we obtain
\be
{{\cal N}^{-1}} - 3 = {\frac 6 {{Z_+}{Z_-}}} (Z - {Z_+})(Z-{Z_-}).
\label{Zquad2}
\ee
It then follows from eqns. (\ref{dtSig1}), (\ref{Zdef}) and (\ref{Zquad2}) that $Z$ satisfies the equation of motion
\be
{\partial _t} Z = {\frac {(Z-{Z_+})(Z-{Z_-})Z} {{\textstyle {\frac 1 2}} {Z_+}{Z_-} + (Z-{Z_+})(Z-{Z_-})}}.
\label{dtZ}
\ee
From eqn. (\ref{dtZ}) we immediately obtain the following qualitative picture of spike formation: suppose that at the initial time there
is a region where $N_1$ is positive and a region where $N_1$ is negative.  Then by continuity there must be a surface 
where $N_1$ vanishes.  It then follows from eqn. (\ref{dtN1}) that on this surface $N_1$ will always be zero, and it then follows from 
eqns. (\ref{bklNeqn2}) and (\ref{Zquad2}) that $Z={Z_-}$ on this surface.  Now consider a point near this surface.  Then $N_1$ is small but nonzero, and therefore $Z$ is close to, but not equal to $Z_-$.  It then follows from 
eqn. (\ref{dtZ}) that the evolution takes $Z$ from near 
$Z_-$ at the initial time to asymptotically close to $Z_+$ at large negative time (recall that the convention is that 
$ t \to - \infty$ as the singularity is approached).  Thus, the surface ${N_1} =0$ is stuck in the old phase, while all nearby
points eventually bounce to the new phase.  Thus a feature of ever more narrow size forms in the vicinity of the surface. 

But we can do even better than this qualitative picture and obtain a complete quantitative picture by integrating eqn. (\ref{dtZ}).  Suppose that at some spatial point at time $t_0$ we have $Z=Z({t_0})$.  Then some straightforward but tedious algebra leads to the following integral of eqn. (\ref{dtZ}):
\bea
\exp& & \left [ {\frac 2  {Z_+}} ({Z_+}-{Z_-})({t_0}-t) \right ] 
\nonumber
\\
&\ &= \left ( {\frac {Z-{Z_-}}  {Z({t_0}) - {Z_-}}}\right ) 
{{\left ( {\frac {{Z_+} - Z } {{Z_+}-Z({t_0})}}\right ) }^{-{Z_-}/{Z_+}}} \times \nonumber\\
&\ &\ \ \ {{\left ( {\frac Z  {Z({t_0})}}\right ) }^ {- 3 ({Z_+} - {Z_-})/{Z_+}}}.
\label{Zsoln}
\eea
Now consider eqn. (\ref{Zsoln}) in the vicinity of a spike.  Choose time $t_0$ sufficiently early in the process that no sharp features have formed, and choose a local coordinate $x$ to vanish where $N_1$ vanishes.  Then for sufficiently small $x$ we have that 
$N_1$ is well approximated by ${N_1} = \epsilon x$ where $\epsilon$ is a function of the coordinates transverse to $x$.  
It then follows from eqns. (\ref{bklNeqn}) and (\ref{bklNeqn2}) that near $x=0$ we have
\be
Z({t_0}) = {Z_-} \left [ 1 + {\frac {Z_+} {{Z_+}-{Z_-}}} {{\left ( {\frac {\epsilon x} 6}\right ) }^2} \right ].
\label{Z0}
\ee
Then using eqn. (\ref{Z0}) in eqn. (\ref{Zsoln}) we obtain
\bea
& &{{\left ( {\frac {\epsilon x} 6}\right ) }^2} 
\exp \left [ {\frac 2 {Z_+}} ({Z_+}-{Z_-})({t_0}-t) \right ] ,
\label{Zx}
\\
&\ &= \left [ {\frac {{({Z_+}-{Z_-})}^2} {{Z_+}{Z_-}}} \right ] \left [ {\frac {Z - {Z_-}} {{Z_+} - Z}} \right ] 
{{ \left [ {\frac {({Z_+} - Z){Z_- ^3}} {({Z_+}-{Z_-}){Z^3}}} \right ] }^{({Z_+}-{Z_-})/{Z_+}}}.
\nonumber
\eea
Eqn. (\ref{Zx}) shows that spikes are essentially a co-dimension one phenomenon, since everything can be expressed in terms of a single coordinate orthogonal to the spike surface.  Thus one should obtain essentially the same behavior in a 2-dimensional simulation as in a 1-dimensional simulation.

We now consider how to compare the results of the simulations to the prediction of eqn. (\ref{Zx}).  Though so far we have talked about the eigenvalues of ${\Sigma ^\alpha}_\beta$ and ${N^\alpha}_\beta$, all the information about the eigenvalues of a matrix is contained in the invariants of that matrix and it is far simpler to compute invariants than to compute eigenvalues.  In particular, since ${N^\alpha}_\beta$ has only one non-negligible eigenvalue, $N_1$, we
find that the invariant ${N^\alpha}_\alpha$ is simply equal to $N_1$.  It then follows from eqns. (\ref{bklNeqn2}) and 
(\ref{Zquad2}) that
\be
{{N^\alpha}_\alpha} = {\frac {\pm 6} {\sqrt {{Z_+}{Z_-}}}} {{({Z_+}-Z)}^{1/2}} {{(Z-{Z_-})}^{1/2}}.
\label{trN}
\ee
Note that eqns. (\ref{trN}) and (\ref{Zx}) together give a parametric equation for ${N^\alpha}_\alpha$ as a function of $x$ (because the equations give both $x$ and ${N^\alpha}_\alpha$ as functions of $Z$).  Thus to make a comparison with simulations, one should find from the simulation ${N^\alpha}_\alpha$ as a function of $x$ and compare to this parametric curve.  

We now consider the behavior of the invariants of ${\Sigma ^\alpha}_\beta$.  Define the quantity $\cal S$ by
\be
{\cal S} = {{\Sigma ^\alpha}_\beta}{{\Sigma ^\beta}_\gamma}{{\Sigma ^\gamma}_\alpha}.
\label{Sdef}
\ee
Since ${\Sigma ^\alpha}_\beta$ is trace-free, it follows that the invariants of ${\Sigma ^\alpha}_\beta$ are
${\Sigma ^{\alpha \beta}}{\Sigma _{\alpha \beta}}$ and $\cal S$.  However, from eqn. (\ref{bklconstraintG}) and the fact that $N_2$ and $N_3$ are negligible, it follows that 
${\Sigma ^{\alpha \beta}}{\Sigma _{\alpha \beta}} = 6 - {\frac 1 2} {{({{N^\alpha}_\alpha})}^2}$ so there is no information in ${\Sigma ^{\alpha \beta}}{\Sigma _{\alpha \beta}} $ that is not already contained in ${N^\alpha}_\alpha$.  Therefore in characterizing the invariants of ${\Sigma ^\alpha}_\beta$ we can restrict our attention to $\cal S$.
From eqns. (\ref{Zdef}-\ref{Sig2}) we find
\be
{\cal S} = 6 \left [ 1 - {{(Z-3)}^2} \right ] + 3 b (1-b){Z^2}(Z-4).
\label{Sform}
\ee
Eqns. (\ref{Zx}) and (\ref{Sform}) give a parametric equation for $\cal S$ as a function of $x$.  Thus one should find from the simulations $\cal S$ as a function of $x$ and compare to this parametric curve.

The formulas given in eqns. (\ref{Zx}), (\ref{trN}) and (\ref{Sform}) contain two parameters: $b$ and $\epsilon$.  Thus to make comparisons with the simulations, we must specify how to determine these parameters from the simulations.  To determine $b$ it is helpful to recall the definition of the BKL parameter $u$.  Consider a time before the bounce when $N_1$ is negligible and the eigenvalues of ${\Sigma ^\alpha}_\beta$ are approximately constant.  This is a Kasner era, and the Kasner exponents ${p_1}, \, {p_2}$ and $p_3$ are expressed in terms of the corresponding eigenvalues of ${\Sigma ^\alpha}_\beta$ by ${p_i}=(1+{\Sigma _i})/3$. 
The BKL parameter $u$ is defined by
\be
u = {p_3} / {p_2}.
\ee
Note that since ${\Sigma _2} \le {\Sigma _3}$ it follows that ${p_2} \le {p_3}$ and therefore that $u \ge 1$.  
Then it follows from eqns. (\ref{Sig2}), (\ref{Sig3}), (\ref{Zpm}) using some straightforward algebra that
\bea
\label{b_def}
b = {\frac 1 {{u^2} + 1}},
\label{bsoln}
\\
{Z_-} = {\frac {3({u^2} + 1)} {{u^2} + u + 1}}.
\label{Zminussoln}
\eea
Before the bounce we have $Z \approx {Z_-}$.  Let ${\cal S}_-$ denote the value of $\cal S$ before the bounce.  Then using eqns. (\ref{Sform}), (\ref{bsoln}), and (\ref{Zminussoln}) straightforward but tedious algebra yields
\be
{{\cal S}_-} = 6 - {\frac {81 {u^2} {{(u+1)}^2}} {{({u^2} + u + 1)}^3}}.
\label{uS}
\ee   
As long as $-6 < {{\cal S}_-} < 6$, there is a unique $u > 1 $ such that eqn. (\ref{uS}) is satisfied.  Thus to compute the parameter $b$ in the simulations, we simply compute the invariant $\cal S$ before a bounce and then use eqn. (\ref{uS}) to determine $u$, and then use eqn. (\ref{bsoln}) to determine $b$.  

There are two different ways to determine the parameter $\epsilon$. This parameter is defined so that at time $t_0$ we have ${N_1} = \epsilon x$, so we can simply use the definition to read off $\epsilon$ from the properties of $N_1$ at a time before the bounce.  Alternatively, if we wait until a time $t_1$ at which a spike has formed, we can use the properties of the spike to determine $\epsilon$ as follows: From eqn. (\ref{trN}) it follows that the maximum value of 
${N^\alpha}_\alpha$ occurs at $Z=({Z_+}+{Z_-})/2$.  (Note also that this maximum value is 
$3 ({Z_+}-{Z_-})/{\sqrt {{Z_+}{Z_-}}}$, a prediction that can easily be compared to the simulations).  Let $x_m$ be the value of $x$ at which this maximum value of ${N^\alpha}_\alpha$ occurs.  Then it follows from eqn. (\ref{Zx}) that
\bea
& & {{\left ( {\frac {\epsilon {x_m}} 6}\right ) }^2} 
\exp \left [ {\frac 2 {Z_+}} ({Z_+}-{Z_-})({t_0}-{t_1}) \right ] \nonumber\\
&\ &= \left [ {\frac {{({Z_+}-{Z_-})}^2} {{Z_+}{Z_-}}} \right ] 
{{ \left [ {\frac {4{Z_- ^3}} {{({Z_+}+{Z_-})}^3}} \right ] }^{({Z_+}-{Z_-})/{Z_+}}}.
\label{Zxm}
\eea 
Combining eqns. (\ref{Zx}) and (\ref{Zxm}) we obtain
\bea
& &{{\left ( {\frac x {x_m}} \right ) }^2} 
\exp \left [ {\frac 2 {Z_+}} ({Z_+}-{Z_-})({t_1}-t) \right ] \nonumber\\
&\ &
= \left [ {\frac {Z-{Z_-}} {{Z_+}-Z}}\right ] 
{{ \left [ {\frac {{{({Z_+}+{Z_-})}^3}({Z_+}-Z)} {4 ({Z_+}-{Z_-}){Z^3}}} \right ] }^{({Z_+}-{Z_-})/{Z_+}}}
\label{Zx2}
\eea 
Thus eqns. (\ref{Zx2}) and (\ref{trN}) provide a parametric curve for ${N^\alpha}_\alpha$ {\it vs.} $x$, while eqns. (\ref{Zx2}) and (\ref{Sform}) provide such a curve for $\cal S$ {\it vs. } $x$.

\subsection{Ephemeral nature of universal spike behavior}
\label{ephemeral}

The universal spike formulas of the previous section were derived under the assumption that spatial derivatives are negligible.  However, it follows from the spike formulas that spatial derivatives become arbitrarily large. Can a quantity be both arbitrarily large and negligible?  
In the equations of motion, all spatial derivatives appear multiplied by ${E_\alpha}^i$.  Thus, spatial derivatives of a quantity $F$ are negligible in the equations of motion provided that the quantity ${{E_\alpha}^i}{\partial _i}F$ is negligible.  Specifically, we will use the spike formulas to calculate the quantity ${{E_\alpha}^i}{\partial _i}{{N^\beta}_\beta}$ at the center of the spike.  Let the subscript $I$ denote tetrad component in the direction of the Ith eigenvector of ${\Sigma ^\alpha}_\beta$.  Then using eqns. (\ref{dtE}) and (\ref{dtN1}) and the fact that 
${\cal N} = 1/3$ in the center of the spike we find that
\begin{equation}
{\partial _t} ({{E_I}^i}{\partial _i}{N_1}) = {\frac 1 3} [4 + 2 {\Sigma _1} - {\Sigma _I}] ({{E_I}^i}{\partial _i}{N_1}).
\end{equation}  
It then follows that the magnitude of ${{E_I}^i}{\partial _i}{N_1}$ gets smaller as the singularity is approached if and only if the quantity in square brackets is positive. However, using eqns. (\ref{Zdef}-\ref{Sig3}) and (\ref{bsoln}-\ref{Zminussoln}) we find
\begin{eqnarray}
{\frac 1 3} [4 + 2 {\Sigma _1} - {\Sigma _1}] = {\frac {{u^2} +1} {{u^2} + u + 1}},
\nonumber
\\
{\frac 1 3} [4 + 2 {\Sigma _1} - {\Sigma _2}] = {\frac {u(u-2)} {{u^2} + u + 1}},
\nonumber
\\
{\frac 1 3} [4 + 2 {\Sigma _1} - {\Sigma _3}] = {\frac {2(1-u)} {{u^2} + u + 1}}.
\end{eqnarray}
The first of these quantities is always positive, the last is always negative, and the one in the middle is positive when $u>2$.  What is going on is the following: during this particular epoch, as the singularity is approached, ${E_1}^i$ and ${E_2}^i$ are getting smaller, while ${E_3}^i$ is getting larger.  Since the spatial derivative of $N_1$ is getting larger, it follows that the product of that spatial derivative and ${E_3}^i$ is always getting larger, though if at the beginning of the epoch ${E_3}^i$ starts out very small, it may take some time before this product is non-negligible.  In contrast, ${E_1}^i$ is getting small faster than the spatial derivative of $N_1$ is getting large, so the product of these two quantities is always negligible.  ${E_2}^i$ is getting small at a rate that is faster (resp. slower) than the spatial derivative of $N_1$ is getting large if $u > 2$ (resp. $u < 2$), thus the product of the two quantities may be getting larger or smaller, depending on the value of $u$.  

This line of reasoning suggests that spikes are ephemeral, or at least that there is only a limited time period under which each spike is accurately described by the spike formulas of the previous section.
This is consistent with the transient spike solutions for Gowdy spacetimes
found in~\cite{lim}.

\subsection{Early spikes and late spikes}
\label{earlylate}

We now consider the extent to which we should expect the approximate formulas of section \ref{spikeformula} to match an actual evolution of the Einstein field equations.  The results of section \ref{spikeformula} are based on the assumptions of that section, namely that ${E_\alpha}^i$ and $A_\alpha$ are negligibly small.  We expect this assumption to be better and better satisfied the closer we are to the singularity, {\it i.e.} the longer the simulation is run.  Or to put it another way: we expect late spikes (the ones that occur later in the simulation) to be better modeled by the analytic formulas of section \ref{spikeformula}.  However, a simulation can only be run as long as it maintains enough resolution for accurate results.  Since spikes are features that become very narrow, that means that eventually in every simulation some spike will become sufficiently narrow to make the simulation lose resolution.  Or to put it another way: simulations can only see early spikes.  Thus there is something of a mismatch between the needs of the simulations and the needs of the spike formula: we expect the spike formula to be a crude, rather than exact, model for the early spikes produced in the simulations.   

\section{1D simulations}
\label{1d}

Our methods for the 1 dimensional ({\it i.e.} two Killing field) case are essentially those of ~\cite{ekpyro,ijjas_et_al} but without the scalar field matter.  In particular, we must choose initial data that satisfies the constraint equations, 
eqns. (\ref{constraintCOM}-\ref{constraintG}).  We do this using the York method~\cite{York}.  That is, we write the initial data in terms of a freely specifiable piece and an unknown conformal factor which we solve for numerically.  
The initial data are the following:
\begin{eqnarray}
{{E_\alpha}^i} = {H^{-1}}{\psi ^{-2}} {{\delta _\alpha}^i},
\\
{A_\alpha} = - 2 {\psi ^{-1}}{{E_\alpha}^i}{\partial _i}\psi,
\\
{N_{\alpha \beta}} = 0 \; \; \; ,
\\
{\Sigma _{\alpha \beta}} = {\psi ^{-6}} {Z_{\alpha \beta}}.
\end{eqnarray}
Here $\psi $ is the unknown conformal factor and $H$ is a constant.  The constraint equations require that $Z_{ik}$ 
be divergence free: that is ${\partial ^i} {Z_{ik}} = 0$.  In addition, since $\Sigma _{\alpha \beta}$ is trace-free, so is $Z_{ik}$.  We choose the following simple $Z_{ij}$ having both these properties
\begin{equation}\label{id}
{Z_{ik}} = 
\begin{bmatrix}
{b_2} & {\kappa_1} & {\kappa_2} \\
{\kappa_1} & {a_1} \cos x + {b_1} & {a_2} \cos x \\
{\kappa_2} & {a_2} \cos x & - {b_1} - {b_2} - {a_1} \cos x  
\end{bmatrix}.
\end{equation}
Here, ${a_1}, \, {a_2}, \, {b_1}, \, {b_2}, \, {\kappa_1}$ and $\kappa_2$ are constants.  The constraint equations require that the conformal factor satisfy the equation
\begin{equation}
{\partial ^i}{\partial _i} \psi = {\textstyle {\frac 3 4}} {H^2} {\psi ^5} - {\textstyle {\frac 1 8}} 
{Z^{ik}}{Z_{ik}} {H^2}{\psi ^{-7}},
\end{equation}
which we solve numerically.  

\begin{figure*}
\centering
\includegraphics[width=0.95\textwidth]{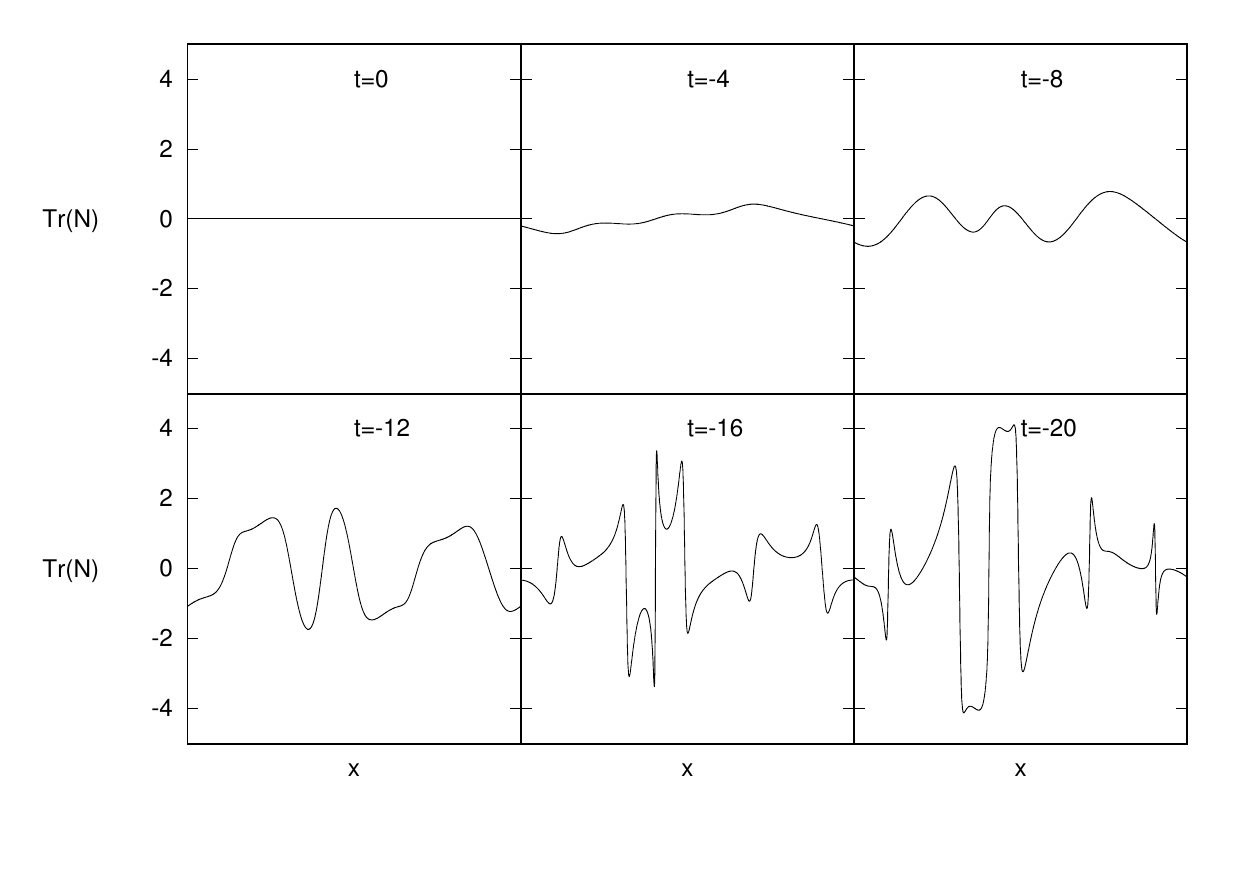}
\caption{$t= {\rm const}$ snapshots of ${N^\alpha}_\alpha$ for $0 \le x \le 2 \pi$ for several different times.  Here the parameters of the initial data (\ref{id}) are ${a_1}=2.5, \, {a_2}=1.2, \, {b_1}=1.5, \, {b_2}=1.2, \, {\kappa_1}=0.4, \, {\kappa_2}=0.3.$ }
\label{fig:1}
\end{figure*}

\begin{figure*}
\centering
\includegraphics[width=0.95\textwidth]{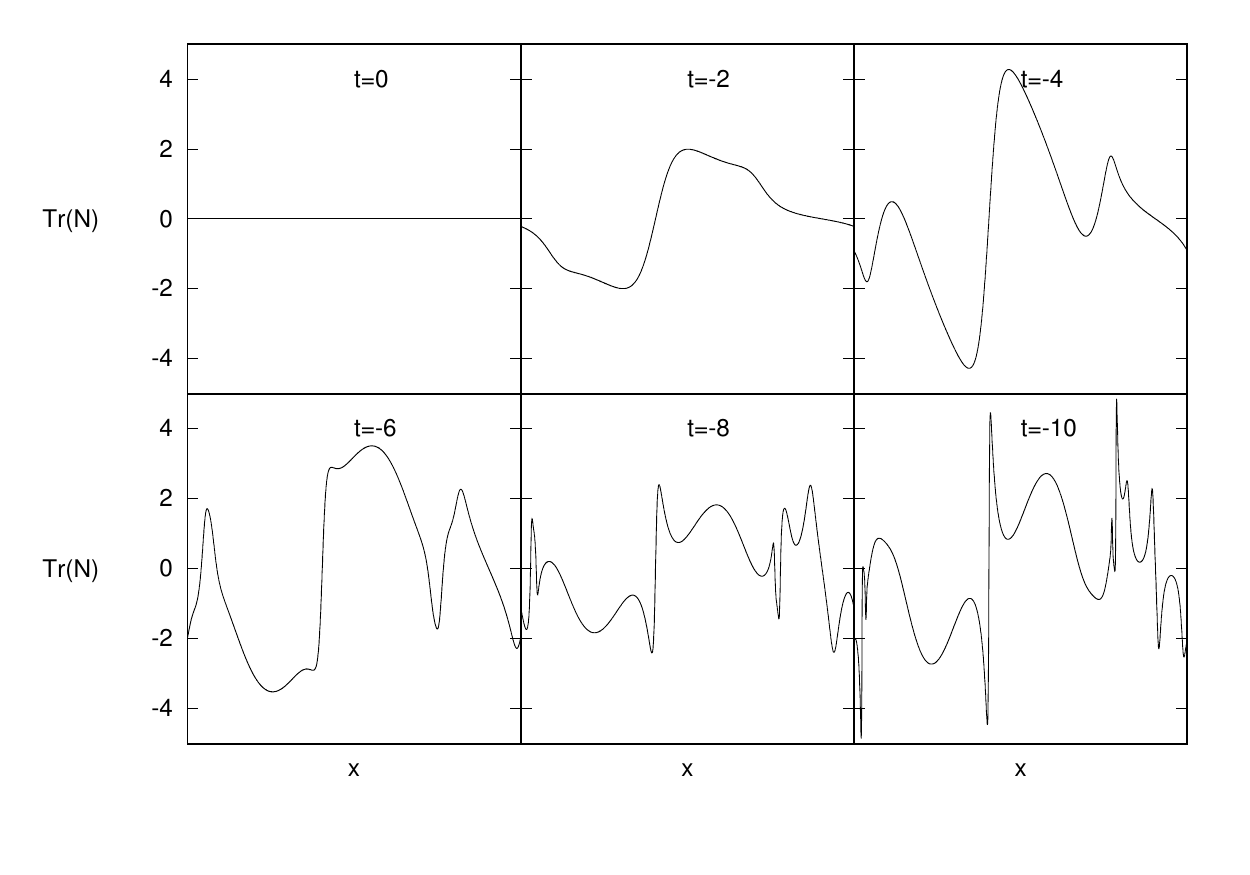}
\caption{$t= {\rm const}$ snapshots of ${N^\alpha}_\alpha$ for $0 \le x \le 2 \pi$ for several different times.  Here the parameters of the initial data (\ref{id}) are ${a_1}=2.0, \, {a_2}=1.2, \, {b_1}=2.0, \, {b_2}=-0.5, \, {\kappa_1}=0.2, \, {\kappa_2}=0.5.$ }
\label{fig:2}
\end{figure*}

\begin{figure*}
\centering
\includegraphics[width=0.95\textwidth]{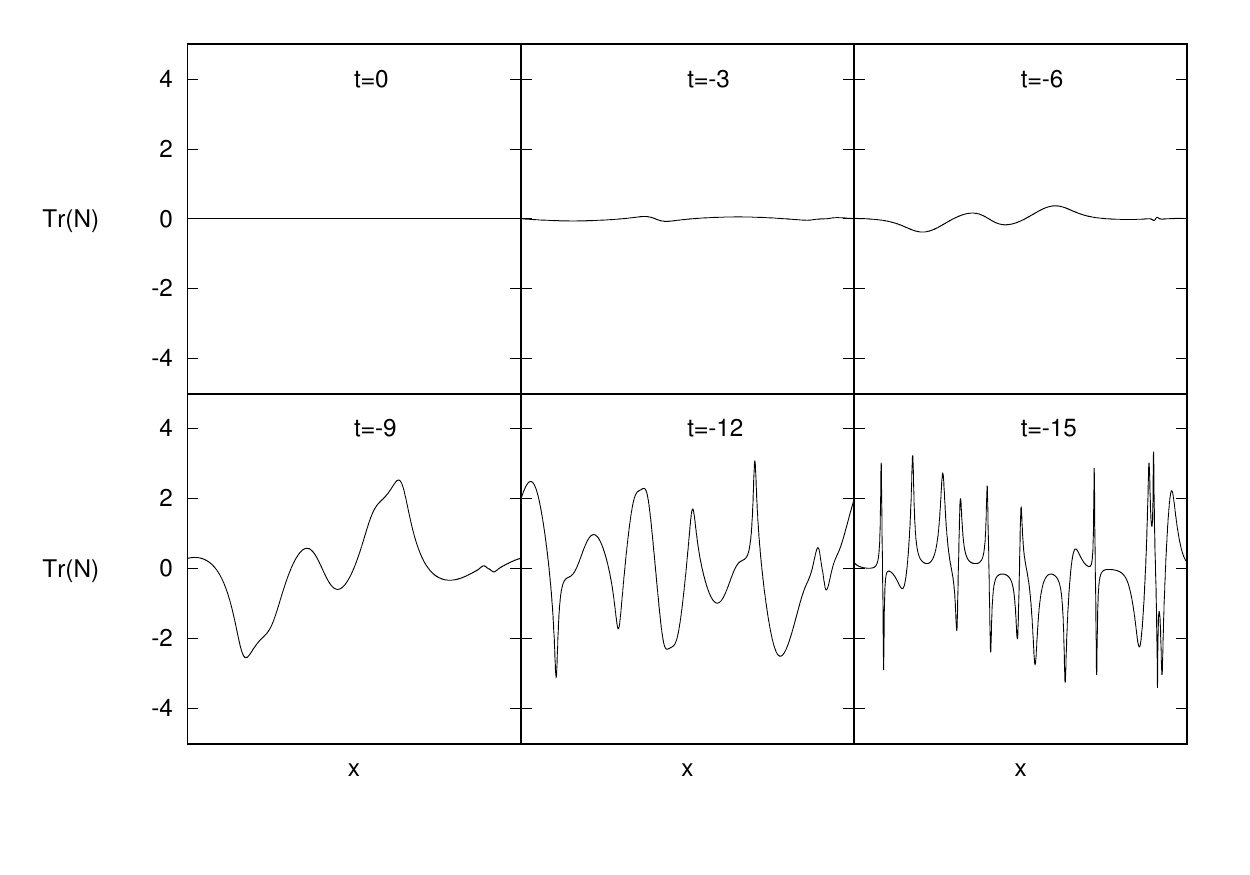}
\caption{$t= {\rm const}$ snapshots of ${N^\alpha}_\alpha$ for $0 \le x \le 2 \pi$ for several different times.  Here the parameters of the initial data (\ref{id}) are ${a_1}=2.5, \, {a_2}=0.5, \, {b_1}=1.0, \, {b_2}=-1.5, \, {\kappa_1}=0.6, \, {\kappa_2}=0.3.$ }
\label{fig:3}
\end{figure*}

\begin{figure}
\centering
\includegraphics[width=0.48\textwidth]{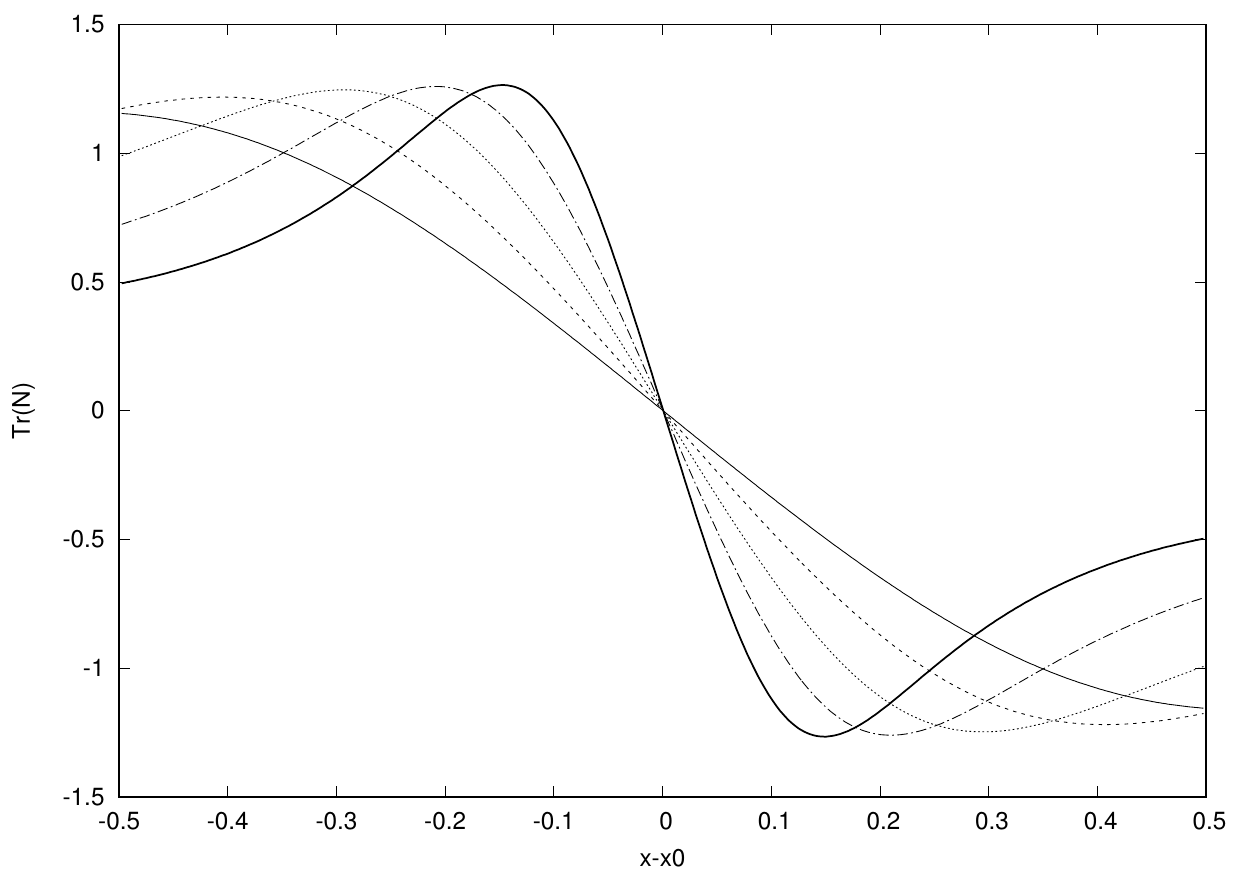}
\caption{${N^\alpha}_\alpha$ vs. $x$ for $t=-12,-13,-14, -15$ and $-16$ for the spike located at 
$x=5.6825$ from the evolution depicted in Fig. \ref{fig:1}.  Here we have translated $x$ so that zero is the center of the spike.} 
\label{Nspike1}
\end{figure}

\begin{figure}
\centering
\includegraphics[width=0.48\textwidth]{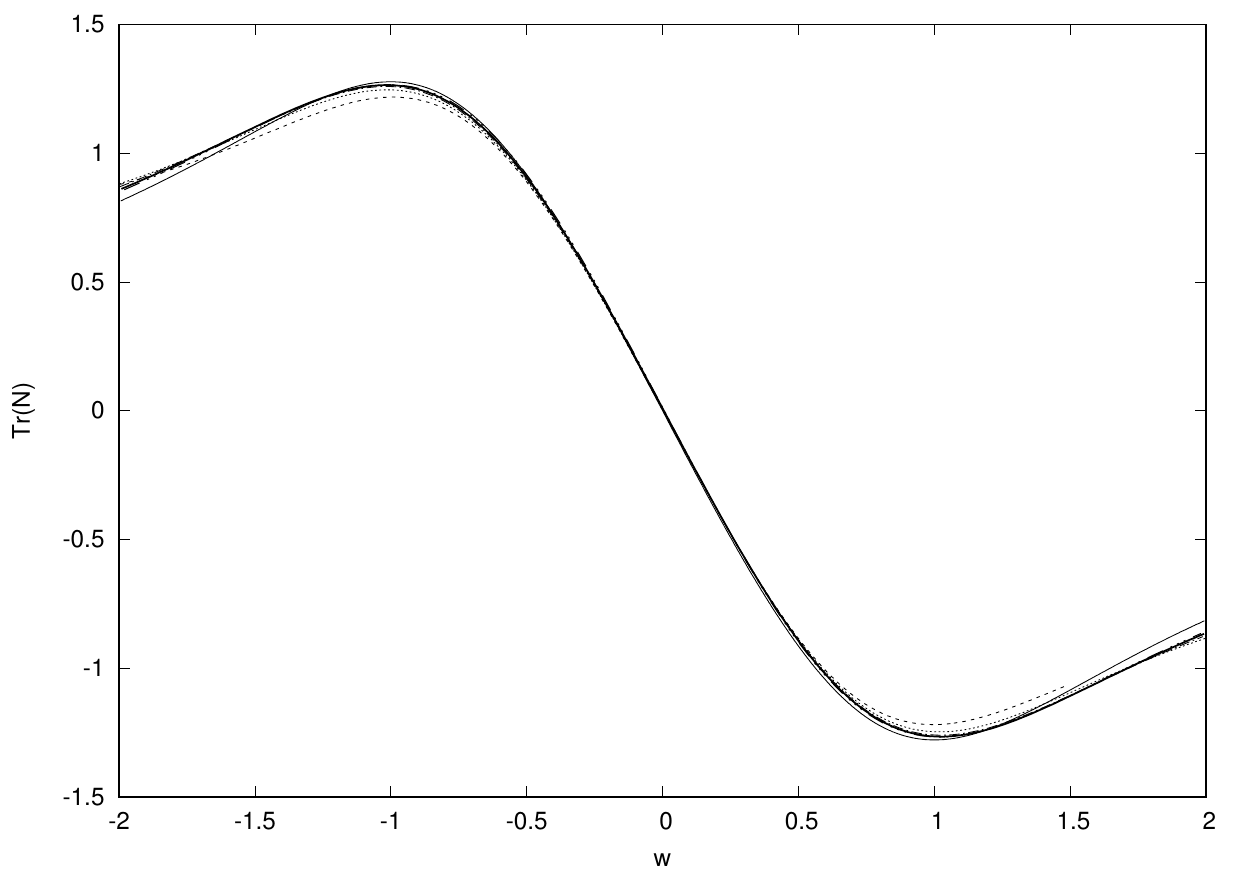}
\caption{${N^\alpha}_\alpha$ vs. the rescaled coordinate $w$ for $t=-12,-13,-14, -15$ and $-16$ for the
same data depicted in Fig. \ref{Nspike1}, along with the spike formula. }
\label{SpikeScale1}
\end{figure}

\begin{figure}
\centering
\includegraphics[width=0.48\textwidth]{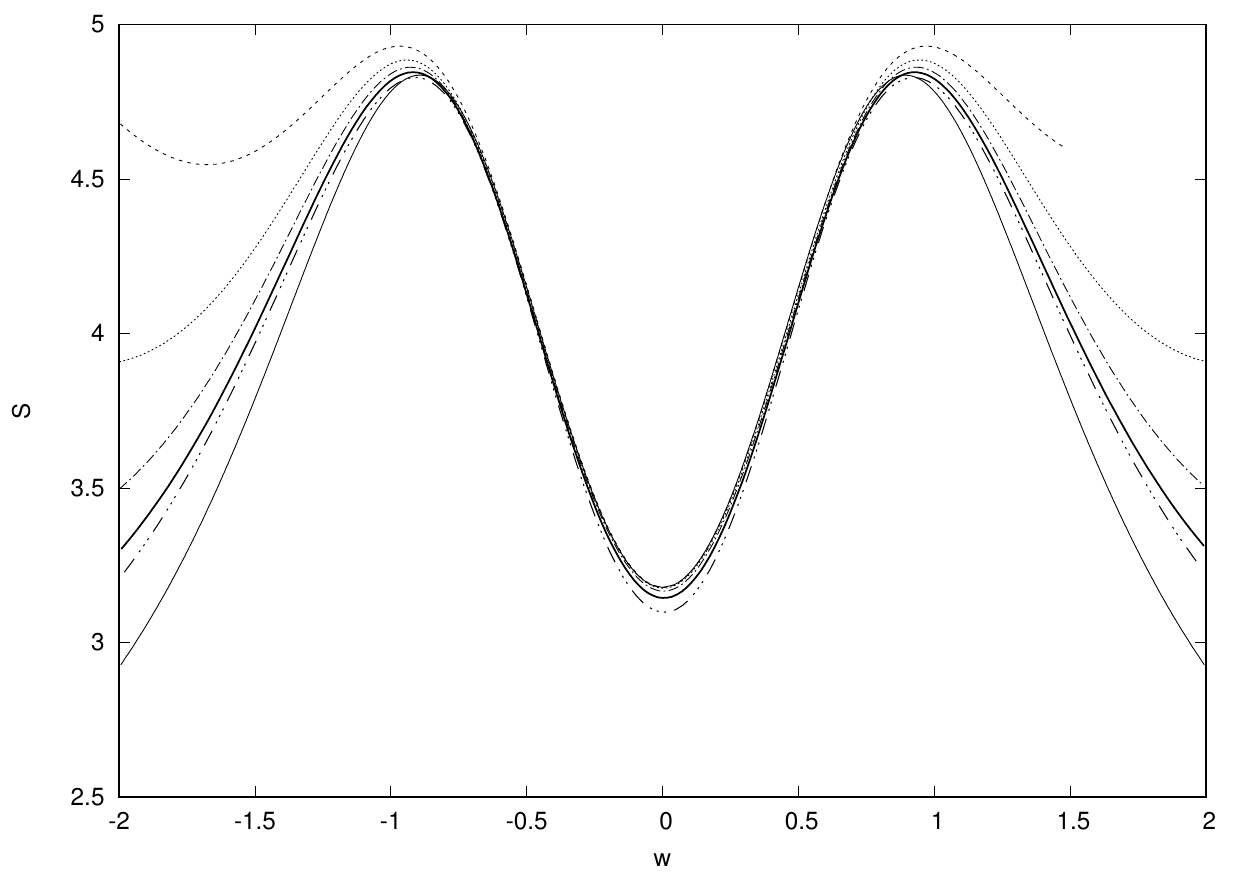}
\caption{$\cal S$ vs. the rescaled coordinate $w$ for $t=-12,-13,-14, -15$ and $-16$, along with the spike formula, for the spike at $x=5.6825$ shown in Fig. \ref{fig:1}. }
\label{S1}
\end{figure}

\begin{figure}
\centering
\includegraphics[width=0.48\textwidth]{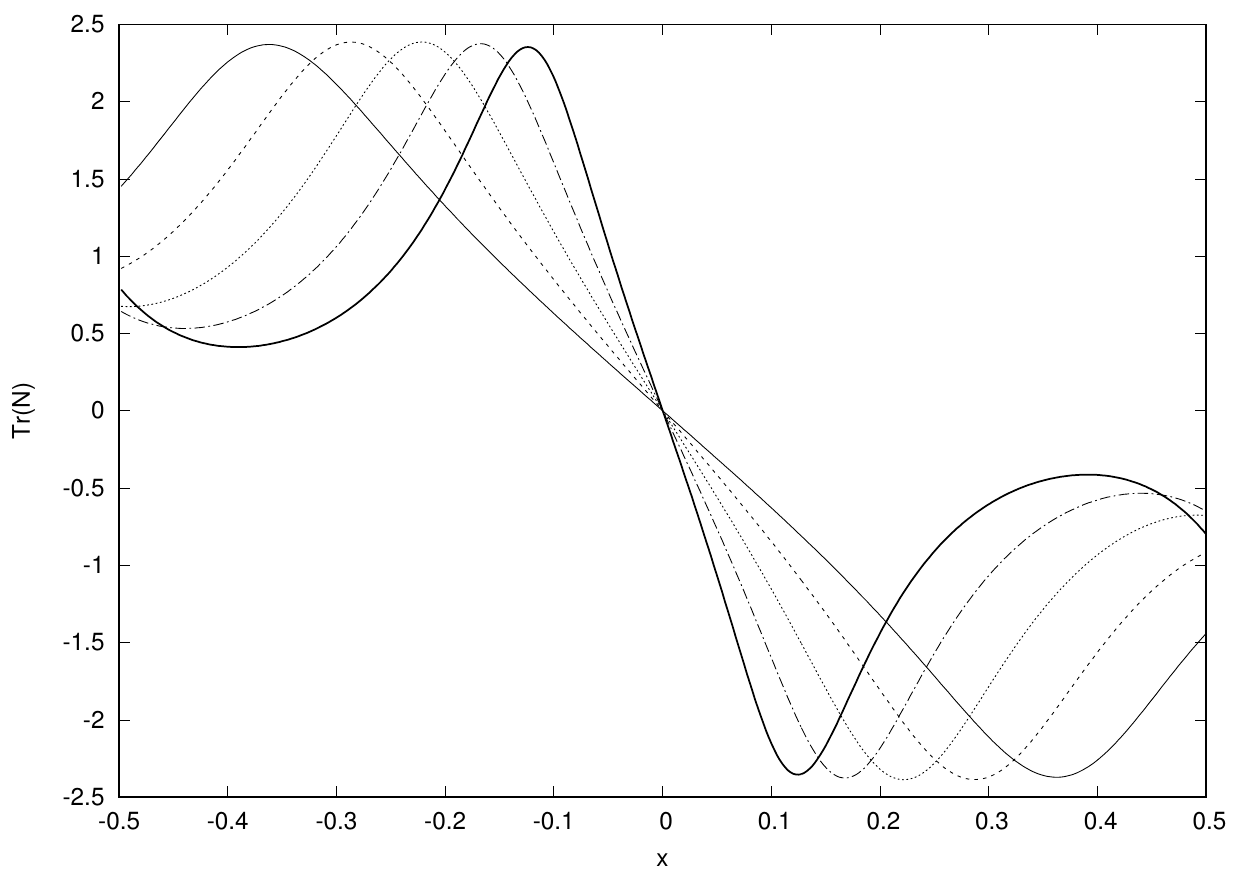}
\caption{${N^\alpha}_\alpha$ vs. $x$ for $t=-7,-7.5,-8, -8.5$ and $-9$ for the spike located at $x=5.683$
from the evolution depicted in Fig. \ref{fig:2}. Here we have translated $x$ so that zero is the center of the spike. }
\label{Nspike2}
\end{figure}

\begin{figure}
\centering
\includegraphics[width=0.48\textwidth]{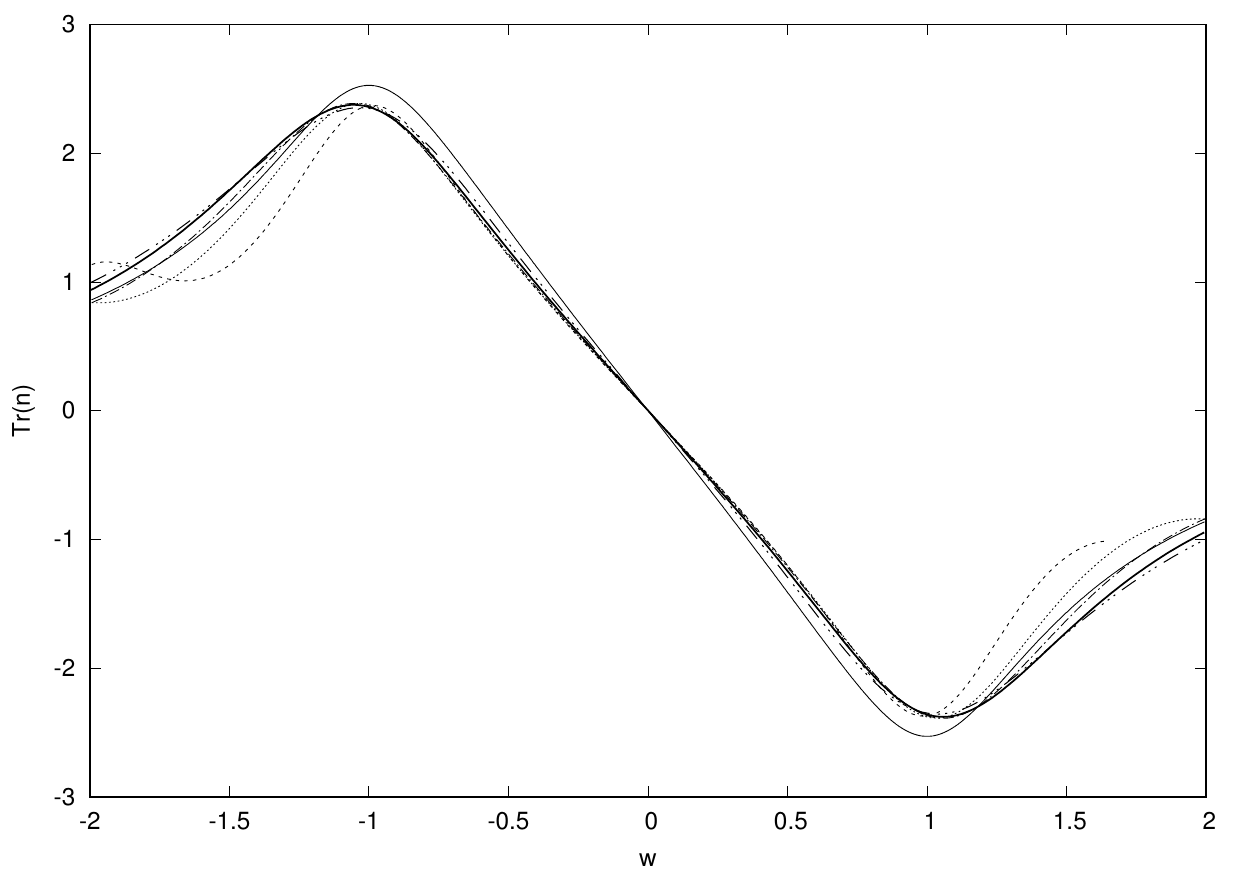}
\caption{${N^\alpha}_\alpha$ vs. the rescaled coordinate $w$ for $t=-7,-7.5,-8, -8.5$ and $-9$
for the same data depicted in Fig. \ref{Nspike2}, along with the spike formula. }
\label{SpikeScale2}
\end{figure}

\begin{figure}
\centering
\includegraphics[width=0.48\textwidth]{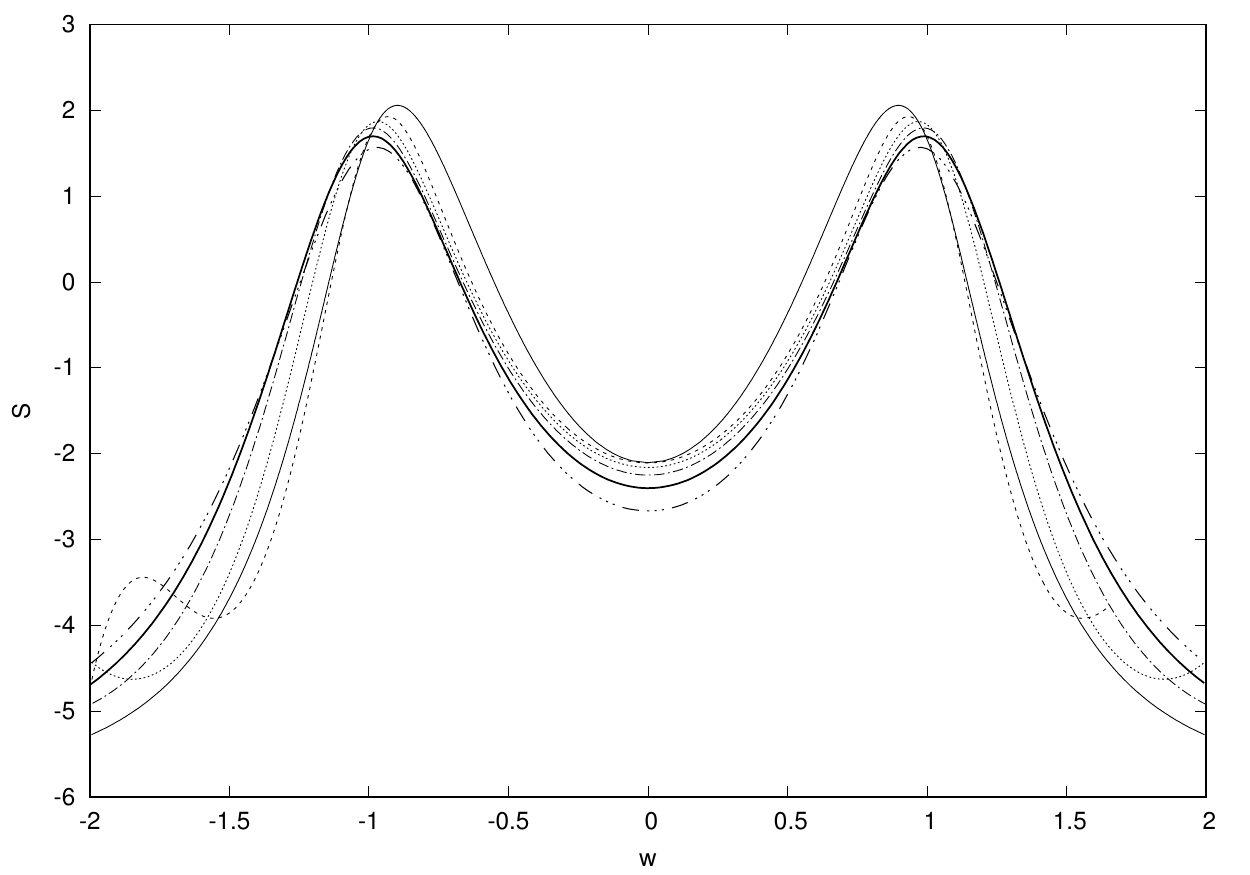}
\caption{$\cal S$ vs. the rescaled coordinate $w$ for $t=-7,-7.5,-8, -8.5$ and $-9$, along with the spike formula, for the spike at $x=5.683$ depicted in Fig. \ref{fig:2}. }
\label{S2}
\end{figure}

\begin{figure}
\centering
\includegraphics[width=0.48\textwidth]{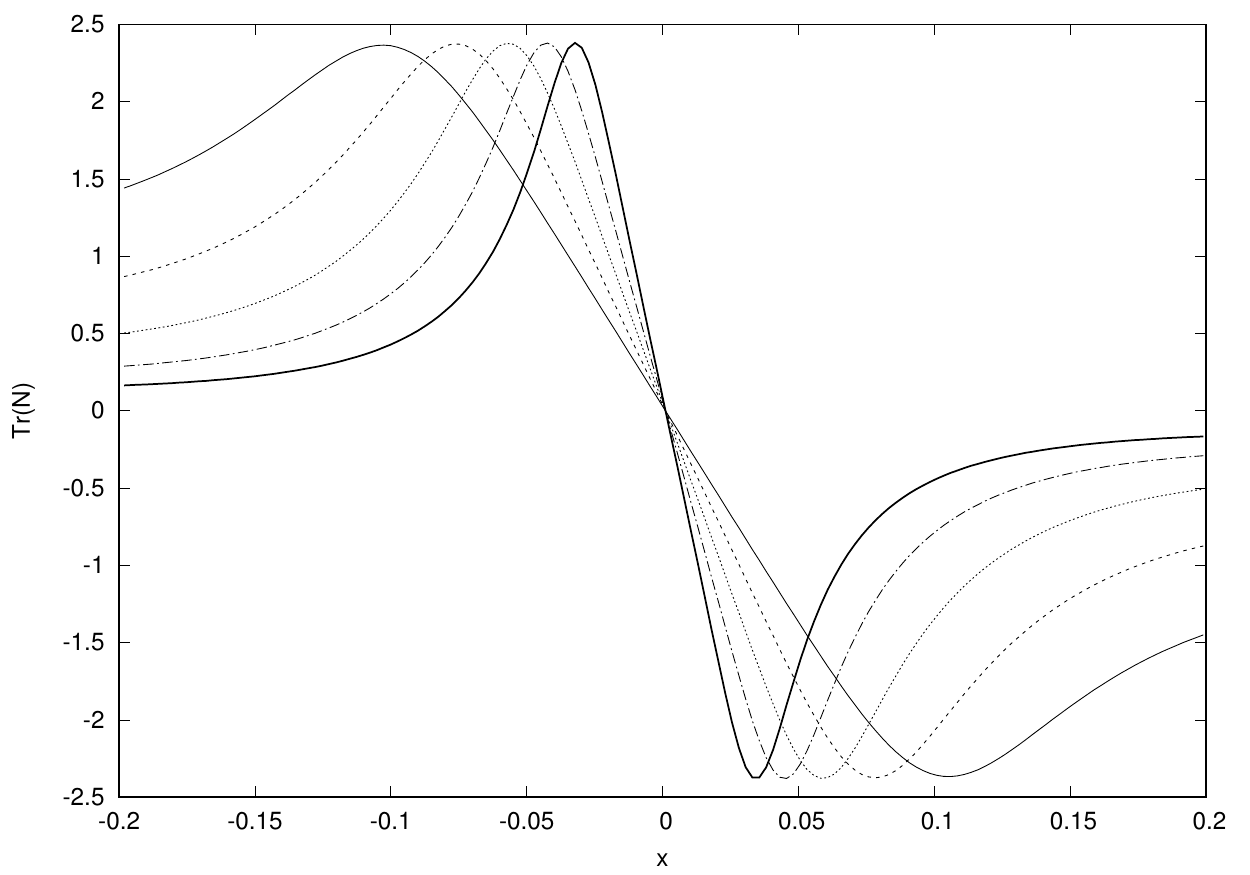}
\caption{${N^\alpha}_\alpha$ vs. $x$ for $t=-13, -13.5, -14, -14.5$ and $-15$ for the spike located at $x=2.5405$ from the evolution depicted in Fig. \ref{fig:3}. Here we have translated $x$ so that zero is the center of the spike.}
\label{Nspike3}
\end{figure}

\begin{figure}
\centering
\includegraphics[width=0.48\textwidth]{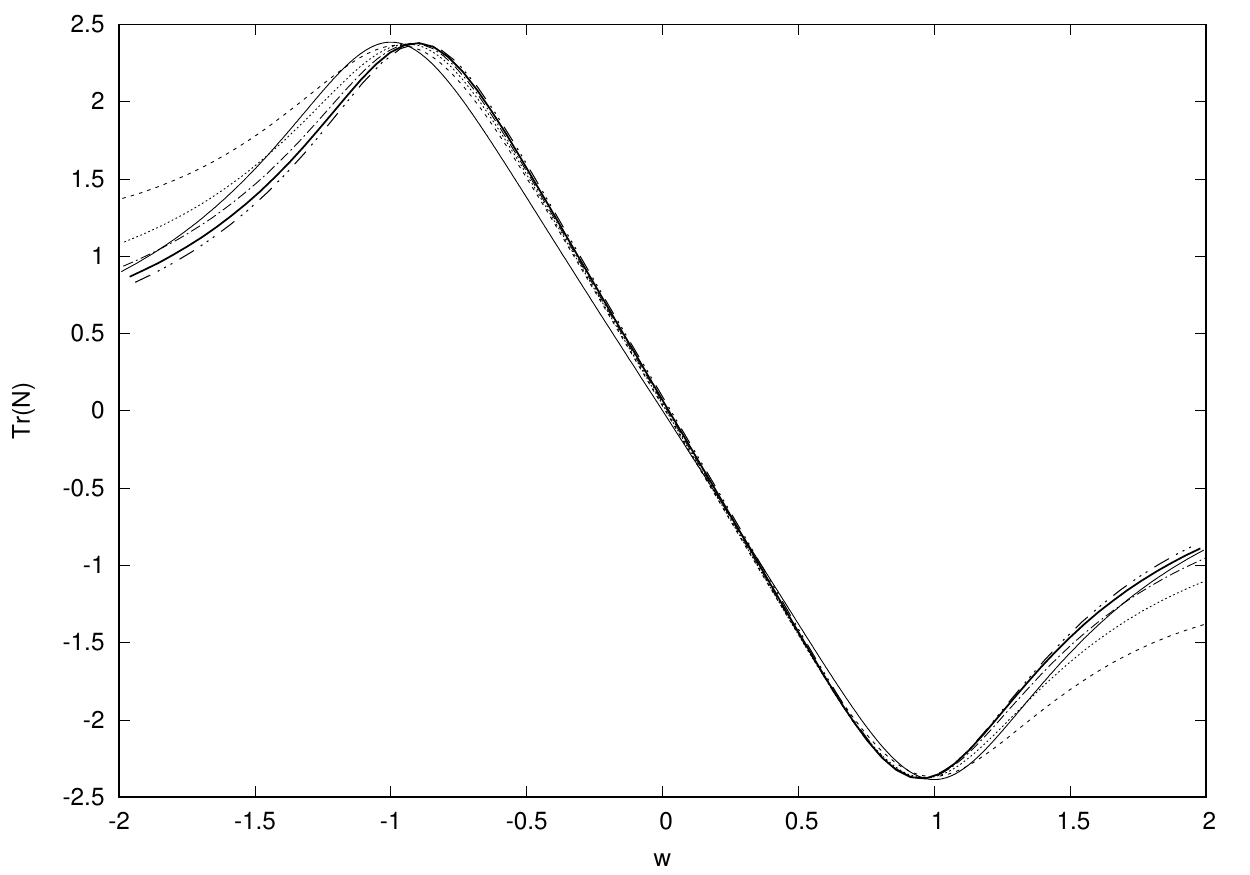}
\caption{${N^\alpha}_\alpha$ vs. the rescaled coordinate $w$ for $t=-13, -13.5, -14, -14.5$ and $-15$, for the same data depicted in Fig. \ref{Nspike3}, along with the spike formula.}
\label{SpikeScale3}
\end{figure}

\begin{figure}
\centering
\includegraphics[width=0.48\textwidth]{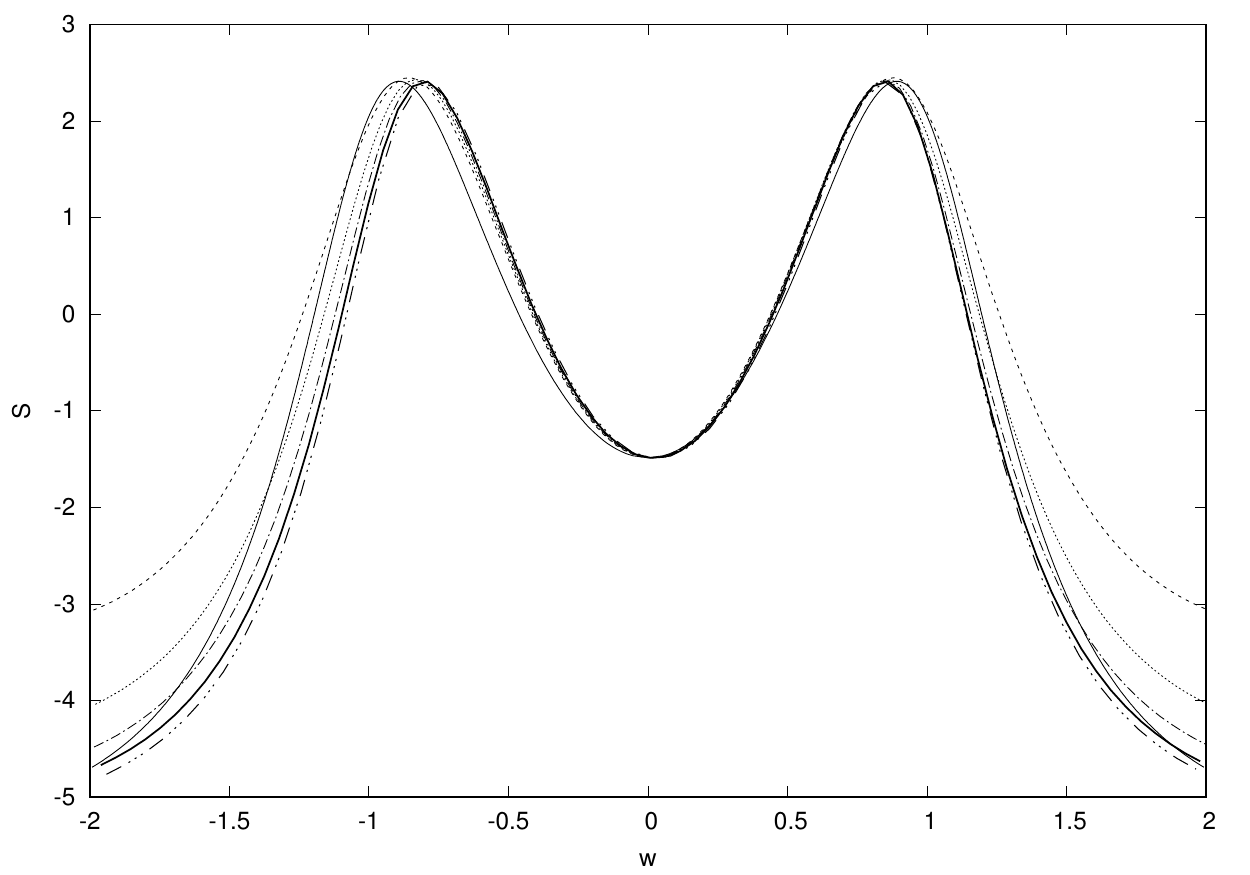}
\caption{$\cal S$ vs. the rescaled coordinate $w$ for $t=-13, -13.5, -14, -14.5$ and $-15$, along with the spike formula, for the spike at $x=2.5405$ depicted in Fig. \ref{fig:3}.}
\label{S3}
\end{figure}

Figures \ref{fig:1}-\ref{fig:3} show the results of three simulations for three different choices of the parameters $({a_1},{a_2},{b_1},{b_2},{\kappa_1},{\kappa_2})$.  In each case snapshots of ${N^\alpha} _\alpha$ vs. $x$ are plotted at different times in the simulation.  The simulations are run with a spatial stepsize of $dx = 2 \pi /2500 \approx 0.00251$ and each simulation is run only for as long as good resolution  can be maintained.  It is clear from the figures that each simulation produces several spikes.  However, as argued in the previous section, the early spikes of a simulation cannot be expected to be well described by the formulas of that section, and even the ``late'' spikes of a simulation are sufficiently ``early'' that the formulas of section \ref{spikeformula} can only be expected to be a fairly crude approximation.  For this reason, we will examine one late spike per simulation.

Figure \ref{Nspike1} shows the spike of the simulation of Fig. \ref{fig:1} that forms at $x= 5.6254$.  In the figure, we have translated $x$ so that the spike is centered at $x=0$.  The figure displays ${N^\alpha} _\alpha$ as a function of $x$ for the times $t=-12,-13,-14, -15$ and $-16$.  

Figure \ref{Nspike1} clearly shows a narrowing feature.  However, to compare with the formulas of section \ref{spikeformula} we must perform a different type of comparison.  Using eqn. (\ref{Zx2}) we define the rescaled spatial coordinate $w$ by 
\begin{equation}
\label{w_def}
w \equiv \left ( {\frac x {x_m}} \right ) \exp \left [ {\frac 1 {Z_+}} ({Z_+} - {Z_-})({t_1} - t) \right ].
\end{equation}   
Then aside from the detailed shape, eqns. (\ref{Zx2}) and (\ref{trN}) contained the prediction that ${N^\alpha}_\alpha$ plotted as a function of $w$, will give the {\emph {same}} shape regardless of time.  Figure \ref{SpikeScale1} contains such a plot.  Here, six different curves are plotted: the five curves of Fig. \ref{SpikeScale1}, but now as a function of $w$, and a sixth curve given parametrically by eqns. (\ref{Zx2}) and (\ref{trN}).  To obtain the parameters in the analytic formula, we find $u$ from the simulations and choose ${t_1} = -12$ and find $x_m$ for that time.  Figure \ref{S1} contains the corresponding six curves for the quantity $\cal S$.  It is clear from Figs. \ref{SpikeScale1} and \ref{S1} that the formulas of section \ref{spikeformula} are a good, but by no means perfect, match to the results of the simulation. 

Figures \ref{Nspike2}-\ref{S2} do the same thing for the simulation of Fig. \ref{fig:2} that Figs. \ref{Nspike1}-\ref{S1} do for the simulation of Fig. \ref{fig:1}.  That is, in Fig. \ref{Nspike2}, one of the late spikes of the simulation of Fig. \ref{fig:2} is plotted as a function of $x$ for five different times.  In Fig. \ref{SpikeScale2}, that same spike is plotted as a function of the rescaled coordinate $w$ along with the corresponding formula from section \ref{spikeformula}.  In Fig. \ref{S2}, the quantity $\cal S$ for the five times is plotted as a function of $w$ along with its formula.  Correspondingly, Figs. \ref{Nspike3}-\ref{S3} perform the same analysis of one of the late spikes of the simulation of Fig. \ref{fig:3}.

In all cases, we find that the formulas of section \ref{spikeformula} are a good but not perfect fit for the results of the simulations.  This is just what we expect from the analysis of that section, due to the fact that even the ``late'' spikes of our simulations are comparatively ``early'' in the sense of section \ref{spikeformula}.

\section{2D simulations}
\label{2d}

The base of the code used for the 2D results is essentially identical
to the 1D code, except now the fields can vary along
two of the spatial dimensions $x$ and $y$, and corresponding discretizations
in the code are represented as 2D arrays. We compactify on a torus,
identifying $x=0$ ($y=0$) with $x=2\pi$ ($y=2\pi$). The same initial data
procedure is used as well, modifying the ansatz for $Z_{ik}$ to 
\begin{widetext}
\begin{equation}
{Z_{ik}} = 
\begin{bmatrix}
{b_2} + a_y\cos(y+\phi_y) & {\kappa_1} & {\kappa_2} \\
{\kappa_1} & {a_1} \cos(x+\phi_x) + {b_1} & 0 \\
{\kappa_2} & 0 & - {b_1} - {b_2} - {a_1} \cos (x+\phi_x)-a_y\cos(y+\phi_y)  
\end{bmatrix}
\;.
\end{equation}
\end{widetext}
Here, ${a_1}, \, {a_y}, \, {b_1}, \, {b_2}, \, {\phi_x}, \, {\phi_y}, \, {\kappa_1} $ 
and $\kappa_2$ are constants. The 2D simulations are computationally quite
expensive compared to the 1D case, so here we only show results for a single 
set of initial data: 
$a_1=0.2,\ a_2=0.7,\ b_1=1.80,\ b_2=-0.15,\ \phi_x=0.15,\ \phi_y=0.25,\ \kappa_1=0.5$ and $\kappa_2=0.3$ .

\begin{figure}
\centering
\includegraphics[width=0.45\textwidth]{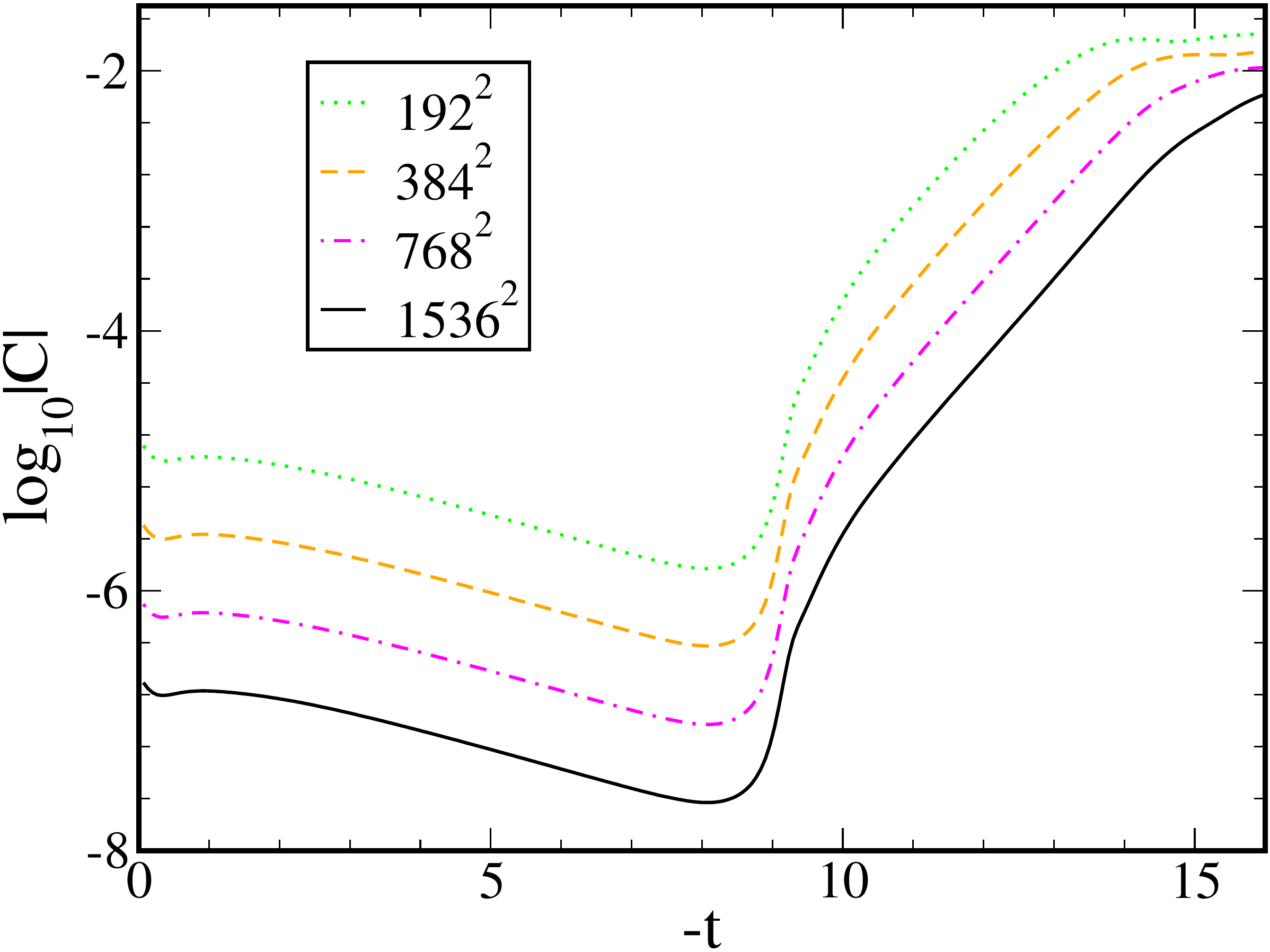}
\caption{An $\ell_2$ norm of all terms in the constraint equations (\ref{constraintCOM}-\ref{constraintG})
over the computational domain, versus time, from 4 different resolution runs of the 2D
case discussed in Sec.\ref{2d}. This
shows close to second order convergence to zero for most of the run time; the drop
in the rate toward the end is due to the spike regions becoming under-resolved.}
\label{conv_2d}
\end{figure}

\begin{figure}
\centering
\includegraphics[width=0.49\textwidth]{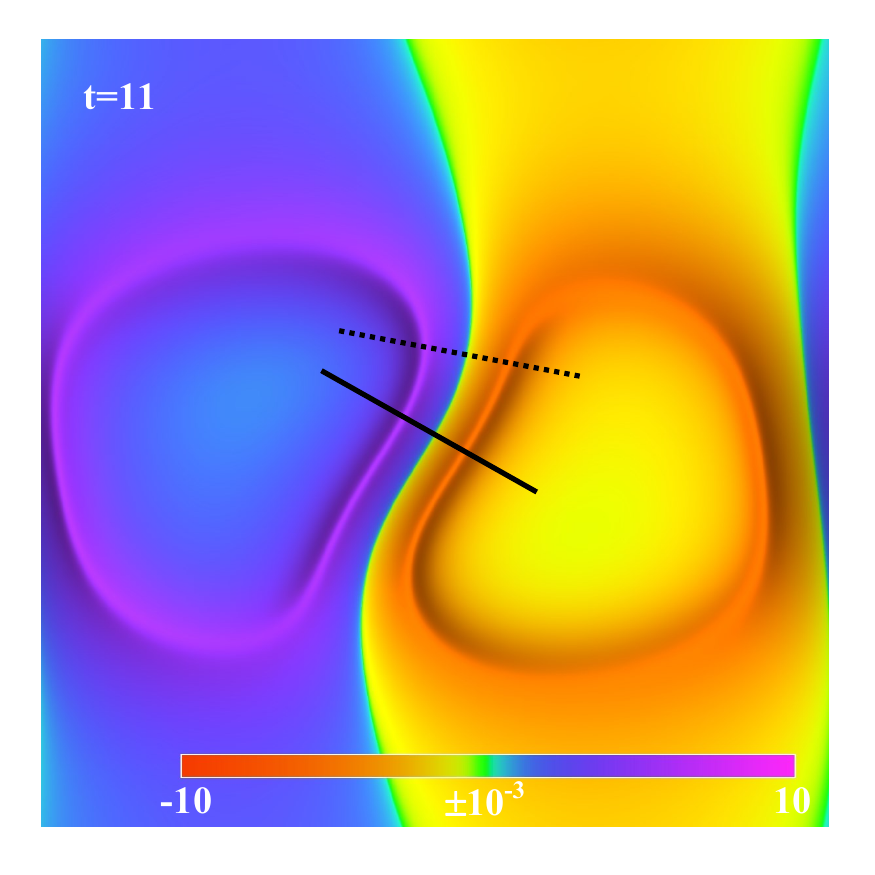}
\includegraphics[width=0.49\textwidth]{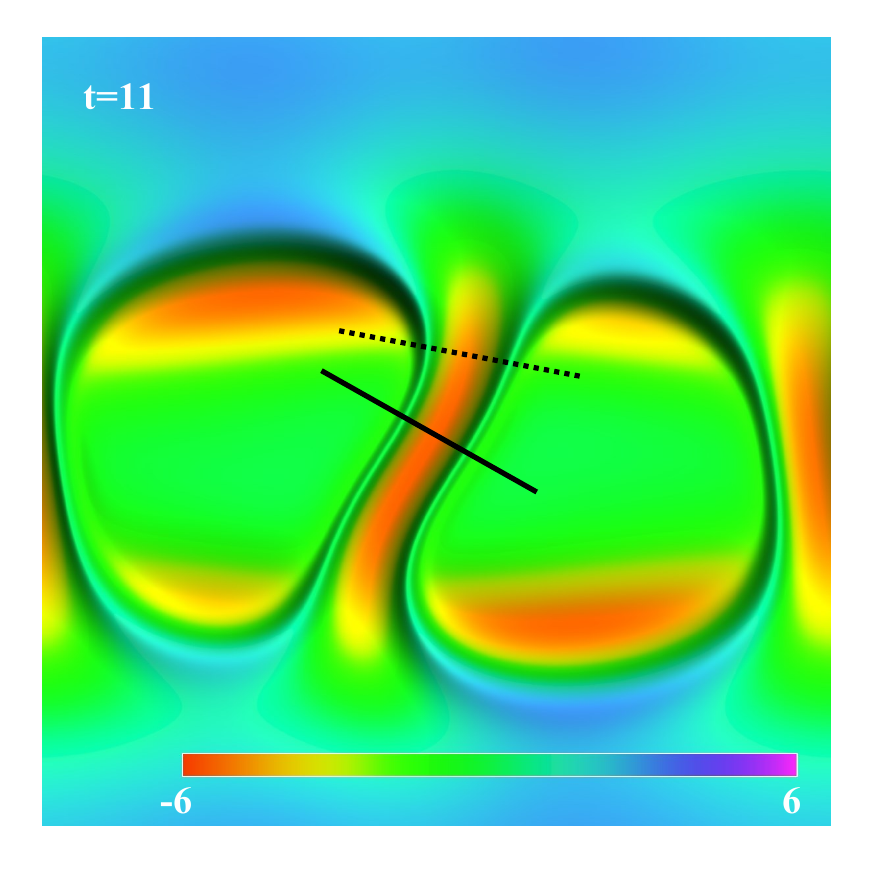}
\caption{Snapshots of $N^\alpha{}_\alpha$ (top) and $\cal S$ (bottom) at $|t|=11$ 
of the 2D simulation. The solid [dashed] line illustrates the slice 
of the domain where the spike profiles centered at $(x,y)=(\pi,\pi)$ [$(x,y)=(3.37,3.75)$],
depicted in Fig. \ref{slices_pi_2d} [Fig. \ref{slices_b_2d}] below, was measured. The width and height
of each picture covers $x=[0,2\pi]$ and $y=[0,2\pi]$ respectively.}
\label{s_n_snapshots_2d}
\end{figure}

As mentioned, the PAMR/AMRD framework allows
for adaptive mesh refinement, however here the spikes are essentially volume filling
(see Fig. \ref{s_n_snapshots_2d}),
and little benefit is achieved compare to unigrid evolution; hence all our runs are unigrid.
To check convergence, the above initial data was evolved
with resolutions $192^2, 384^2, 768^2, 1536^2$; see Fig. ~\ref{conv_2d}
for a plot of the norm of the constraints with time. The comparison
figures shown below were obtained from the highest resolution data.

As discussed, the hypothesis is that spikes form along co-dimension one
volumes of the spacetime where $N^\alpha{}_\alpha=0$. For the 2D case
then, this would correspond to lines within the $(x,y)$ subspace, and the analytic
approximation for the spike profiles should approximate the full (numerical)
results on any slice orthogonal to a given point along the spike line.
The parameters $\epsilon$ and $b$ (see Sec.~\ref{spikeformula}) governing the spike profile can
vary along the spike line. For a given point that we want to compare, we
measure these parameters at one time within the simulation. We find that the
extracted value for $b$, the quantity characterizing the geometry of
the spike point (\ref{b_def}), varies by a few percent depending on what
time we choose to measure it; this is not unexpected, in particular
because we only have the resolution to uncover the early time evolution
of the spike, whereas the analytical formula should govern its late time behavior. 
The parameter $\epsilon$ sets the scale of the spike at a given time, 
so is more a function of the initial data than intrinsic to the spike geometry;
thus we set it to give a best fit to $N^\alpha{}_\alpha$ at the time $b$ is measured.

In the 2D case there is also more gauge ambiguity in performing the comparison
than the 1D case; in particular, how to define ``orthogonal'' far from the spike line,
as well as defining the coordinate measure $w$ (\ref{w_def}) along the spike.
Here we simply define tangent/orthogonal to a spike line as measured in coordinate space
$(x,y)$, setting the overall scale ($\epsilon$) for the orthogonal direction $w$ at the time
the spike parameters are measured, and then assuming the scale narrows with
time as predicted by the analytic formula (i.e., we cannot distinguish between the differences in scale
that arise with time from gauge effects vs limitations
of the approximation).

\begin{figure}
\centering
\includegraphics[width=0.49\textwidth,trim=0in 0in 0in -0.3in,clip]{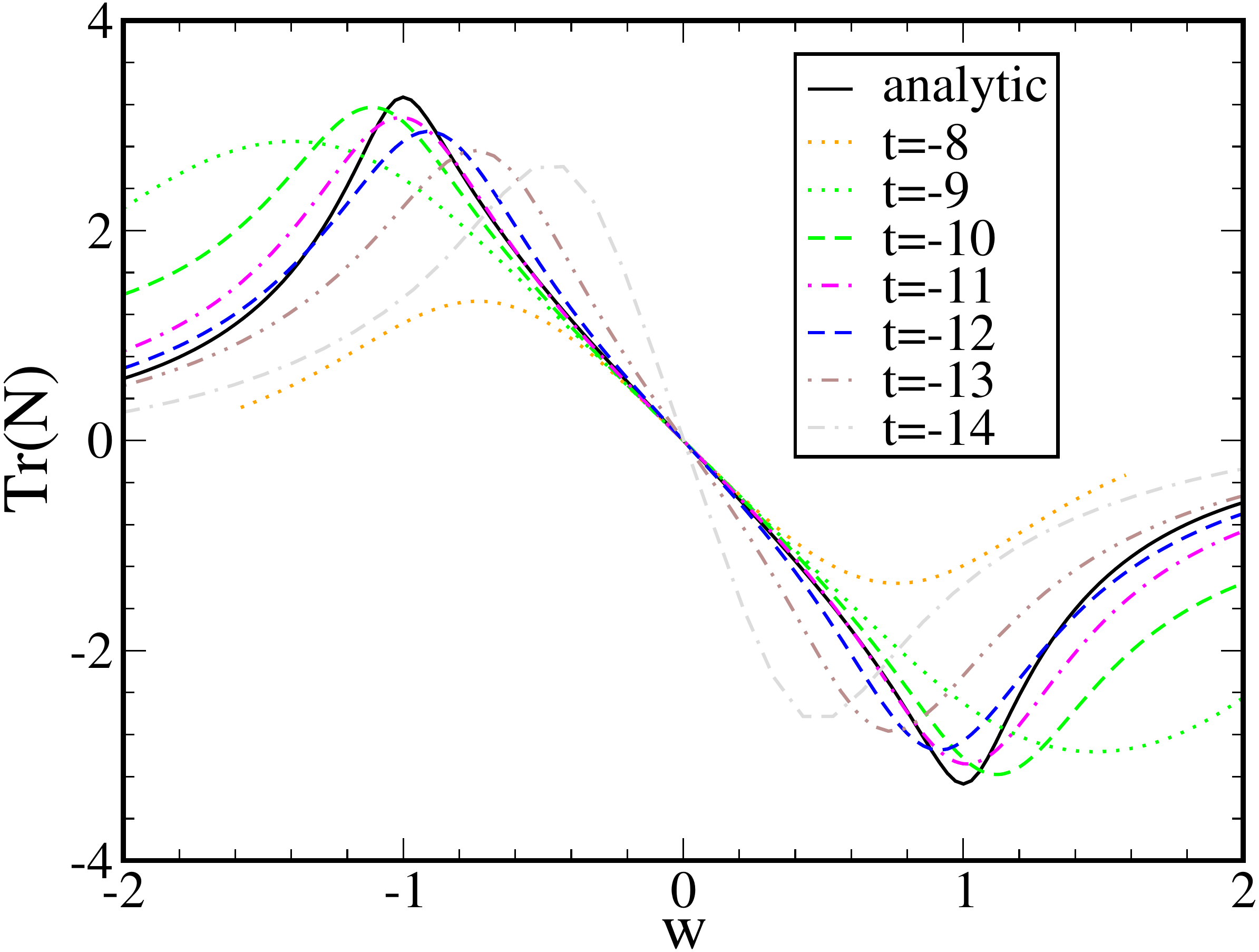}
\includegraphics[width=0.49\textwidth]{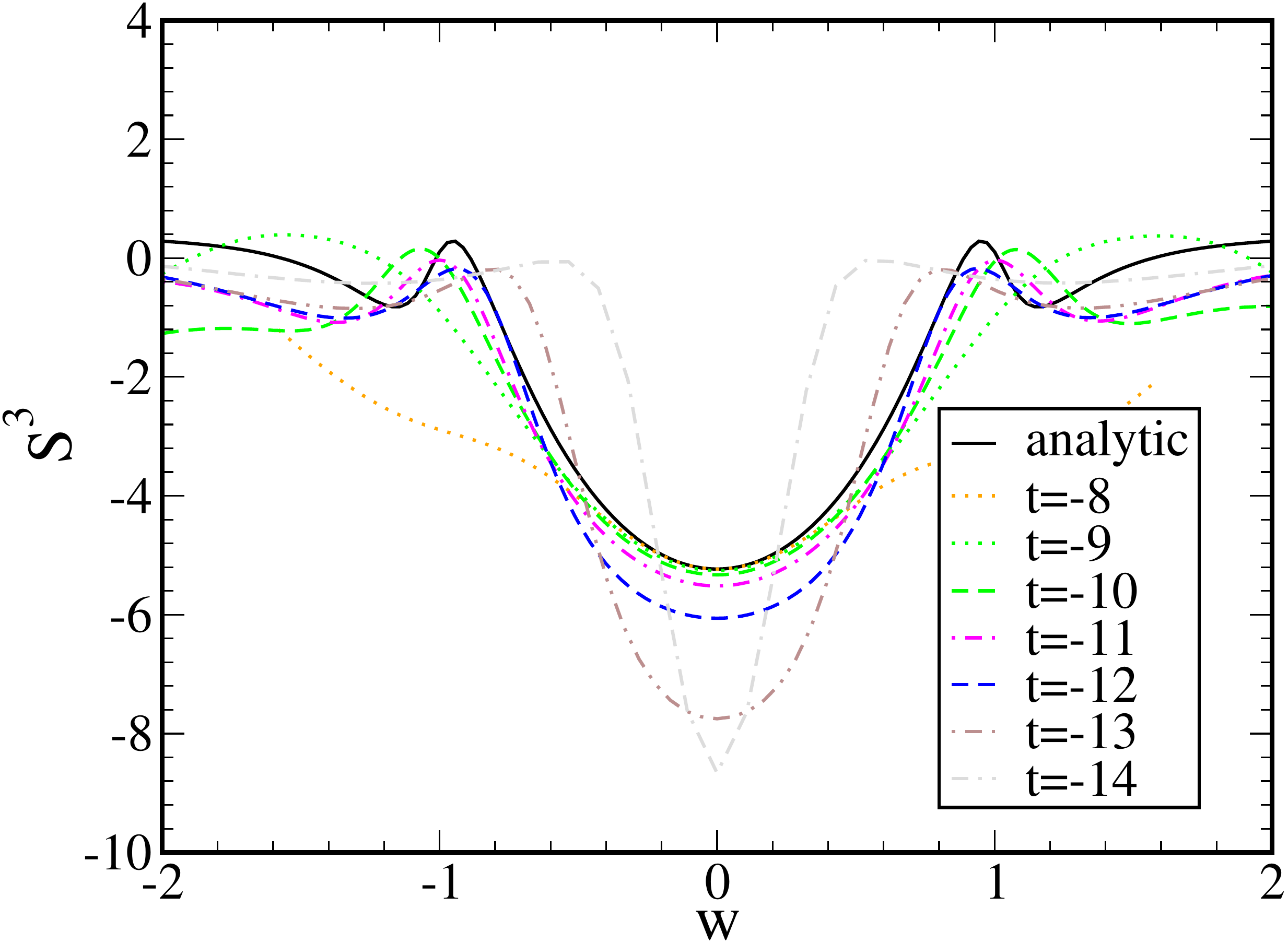}
\caption{$N^\alpha{}_\alpha$ (top) and $\cal S$ (bottom) measured along the slice of the
2D simulation orthogonal to the spike centered at  $(x,y)=(\pi,\pi)$ 
(solid line in Fig. \ref{s_n_snapshots_2d}),
at several times, together with the analytic approximations (for the latter 
the spike parameter $b$ was measured at $t=-8$ to be $b\sim 0.35$).}
\label{slices_pi_2d}
\end{figure}

Here we show a comparison of the numerical results versus analytic formulae
along two slices of the simulation, as depicted in Fig. \ref{s_n_snapshots_2d}.
Figure \ref{slices_pi_2d} shows $N^\alpha{}_\alpha$ and $\cal S$ orthogonal
to a point on the spike line at $(x,y)=(\pi,\pi)$, and Fig. \ref{slices_pi_2d}
for that at $(x,y)=(3.37,3.75)$. 
The results for the 2D runs are thus qualitatively consistent with that
demonstrated for the 1D case : the formulae show decent agreement
at intermediate times of the runs (late enough that a spike has clearly
formed, but not so late that the spike has become under-resolved). 

\begin{figure}
\centering
\includegraphics[width=0.49\textwidth]{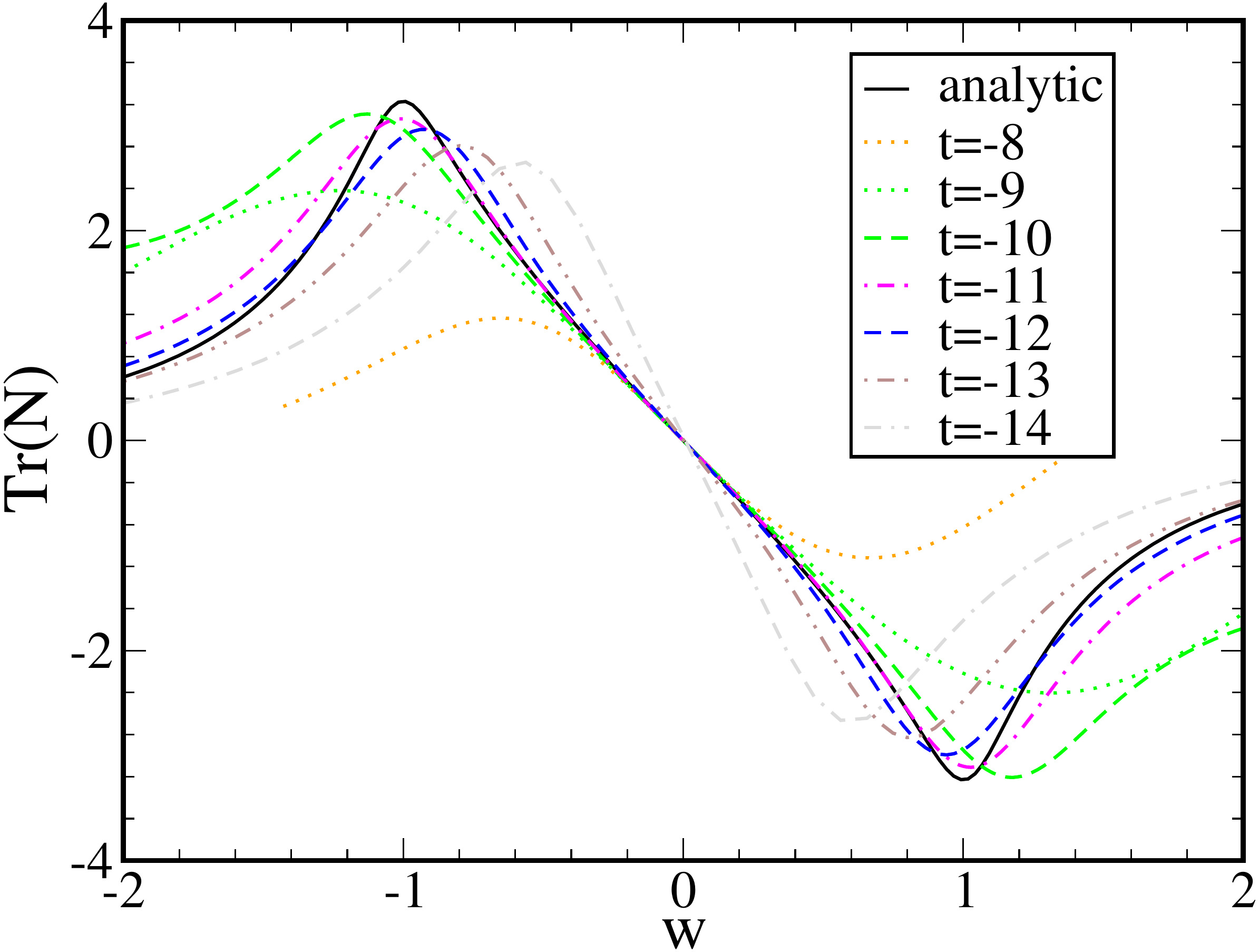}
\includegraphics[width=0.49\textwidth]{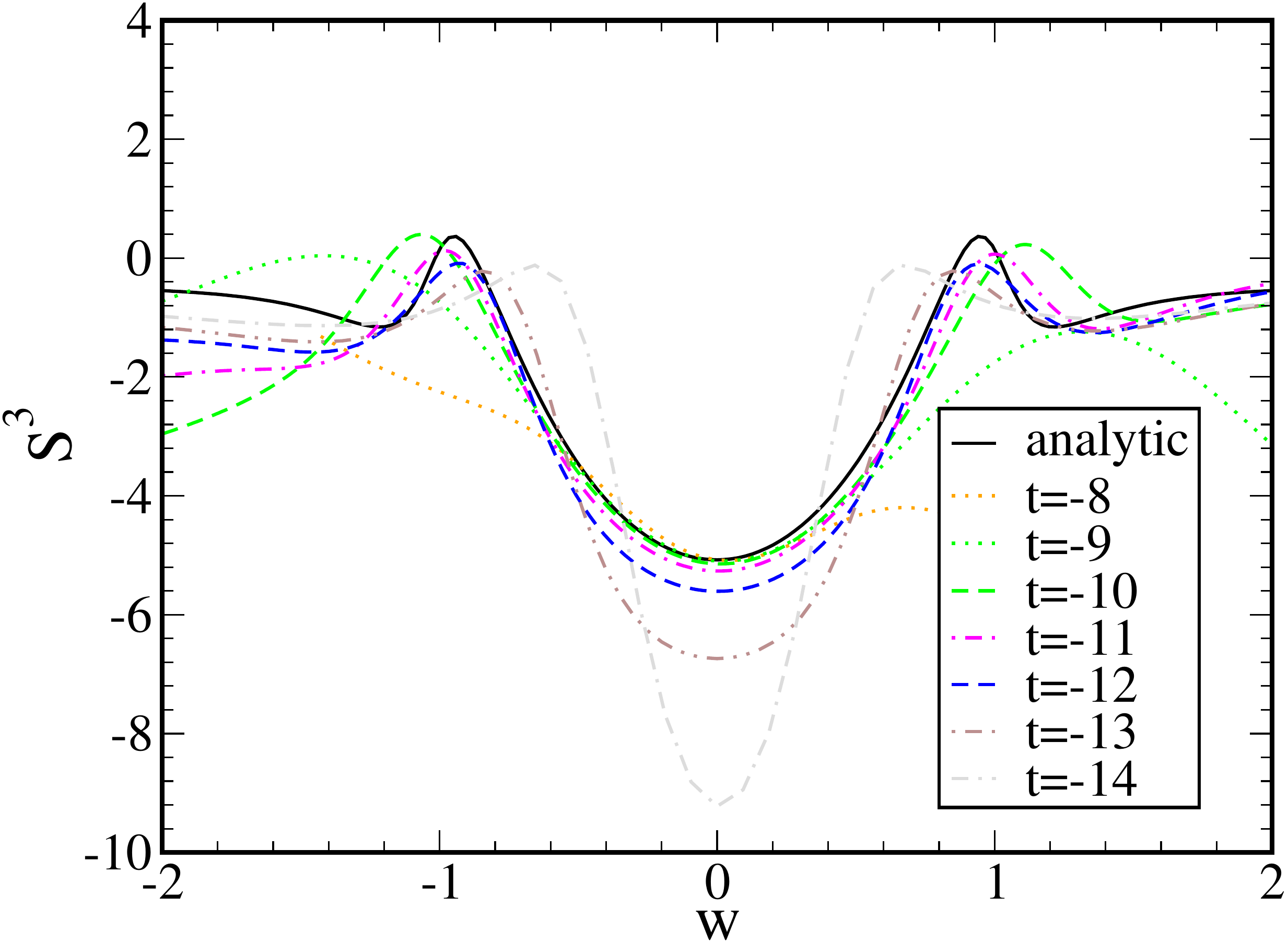}
\caption{$N^\alpha{}_\alpha$ (top) and $\cal S$ (bottom) measured along the slice of the
2D simulation orthogonal to the spike centered at  $(x,y)=(3.37,3.75)$ 
(dashed line in Fig. \ref{s_n_snapshots_2d}),
at several times, together with the analytic approximations (for the latter
the spike parameter $b$ was measured at $t=-8$ to be $b\sim 0.32$).}
\label{slices_b_2d}
\end{figure}

\section{Conclusions}
\label{conclusions}
BKL dynamics consists of a sequence of bounces in the approach to the singularity.  When spikes were first found in the simulations of~\cite{beverlyandvince1} they seemed like a mysterious 
exception to the behavior of the rest of the spacetime.  Instead we see that spikes are a straightforward consequence of BKL behavior.  Each bounce is driven by growth in ${N^\alpha}_\alpha$.
But in general ${N^\alpha}_\alpha$ vanishes on surfaces of co-dimension one.  Points on that surface don't bounce, while nearby points do, leading to an ever narrower feature: the spike.
This qualitative picture gives rise to a quantitative description encapsulated in the formulas of section \ref{spikeformula} for the behavior of the invariants of $N_{\alpha \beta}$ and 
$\Sigma_{\alpha \beta}$ as functions of transverse distance from the spike.  

Spikes are a significant challenge for numerics, due to the need to resolve small scale features at so many points as to make adaptive mesh refinement impractical.  This places severe 
limitations on the amount of time for which such a simulation can be run.  However, the BKL approximation itself (and its consequences like the spike formulas) gets better the closer the 
singularity is approached, and thus the longer the simulation is run.  The simulations of this paper are a compromise between these two stringent requirements: long enough to come within 
the regime of validity of the BKL approximation, but short enough that resolution is not overwhelmed.

Within this uneasy compromise, we find compelling evidence for the picture of section \ref{spikeformula}.  That is, the simulations match the formulas of that section as well as can be 
expected.  This characterization of spikes completes the numerical evidence that BKL behavior describes the approach to the singularity in spacetimes with compact Cauchy surfaces.

\section*{Acknowledgments}
{
This work is supported in part by NSF grants PHY-1505565 and PHY-1806219 (DG), and NSF grants PHY-1607449 and PHY-1912171, the Simons Foundation, and the Canadian Institute for Advanced Research (CIfAR) (FP).  Computational resources were provided by the Perseus cluster at Princeton University.
}

\end{document}